\documentclass[11pt]{article}

\usepackage{amsfonts}
\usepackage{amsbsy} 
\usepackage{epsfig}
\usepackage{latexsym}
\input amssym.def
\input amssym.tex

\def\bequ{\begin{equation}}
\def\eequ{\end{equation}}
\def\barr{\begin{array}}
\def\earr{\end{array}}

\def\ben{\begin{equation}}
\def\een{\end{equation}}
\def\bena{\begin{eqnarray}}
\def\eena{\end{eqnarray}}

\newcommand{\sect}[1]{\setcounter{equation}{0}\section{#1}}

\advance \topmargin by -\headheight
\advance \topmargin by -\headsep
\evensidemargin \oddsidemargin
\advance\hoffset by -3mm  
\advance\voffset by  8mm  

\begin{document}
\hfuzz=100pt
\title{Supersymmetric Rotating Black Holes and Causality Violation} 
\author{\\ 
G W Gibbons\footnote{E-mail:gwg1@damtp.cam.ac.uk} \& C A R Herdeiro\footnote{E-mail:car26@damtp.cam.ac.uk}
\\
\\
Department of Applied Mathematics and Theoretical Physics,
\\ (D.A.M.T.P.)
\\ Cambridge University, 
\\ Silver Street,
\\ Cambridge CB3 9EW,
 \\ U.K.}

\date{June 1999}
\maketitle

\centerline{DAMTP-1999-78}

\begin{abstract} 
The geodesics of the
rotating extreme black hole in five spacetime dimensions found by Breckenridge, Myers, Peet and Vafa 
are Liouville integrable and may be integrated by additively separating the Hamilton-Jacobi equation. This allows us to obtain the St\"ackel-Killing tensor.
We use these facts to give the maximal analytic extension of the spacetime
and discuss some aspects of its causal structure. In particular, we exhibit a `repulson'-like behaviour occuring when there are naked closed timelike curves. In this case we find that the spacetime is geodesically complete (with respect to causal geodesics) and free of singularities. When a partial Cauchy surface exists, we show, by solving the Klein-Gordon equation, that the absorption cross-section for massless waves at small frequencies is given by the area of the hole. At high frequencies a dependence on the angular quantum numbers of the wave develops. We comment on some aspects of `inertial time travel' and argue that such time machines cannot be constructed by spinning up a black hole with no naked closed timelike curves.

\end{abstract}

\newpage

\sect{Introduction}

The positive energy property of states in a supersymmetric theory
excludes tachyons. Tachyons can lead to causal pathologies
such as the ``tachyon telephone" \cite{tt} . One may therefore ask
whether supersymmetry excludes the existence of closed
timelike curves (``CTC's"). The answer is obviously
no. Both flat space with a periodically identified time coordinate
and Anti-de-Sitter spacetime, AdS
(rather than its universal covering spacetime $\widetilde{AdS}$), 
admit CTC's; in fact they admit closed timelike geodesics (``CTG's").
They 
also admit the maximum possible number of Killing spinors.
However, these two examples are not simply connected and the 
CTC's may be eliminated by passing to a 
topologically trivial covering spacetime. 

G\"odel \cite{godel} first
realized\footnote{Causality violating spacetimes obeying Einstein equations were found prior to G\"odel`s work, most notably by Lanczos \cite{lanczos} and Van Stockum \cite{vans}. However, G\"odel was indeed (to the best of our knowledge) the first to exhibit a spacetime with CTC's \textit{and} explicitly discuss the possibility to ''travel into the past``.} that one could have CTC's in a 
completely non-singular, topologically trivial
solution of the Einstein equations with matter satisfying the 
dominant energy condition. However, his solution is neither
asymptotically flat (in fact it is a homogeneous spacetime)
 nor (as far as we are aware) supersymmetric. 

It was only recently that simply connected,
asymptotically flat
BPS solutions of the supergravity equations of 
motion which  admit CTC's were found to exist \cite{BMPV,kallosh}. 
These spacetimes, which we refer to as the `BMPV family', have a net angular momentum
and  represent, for sufficiently small
values of the angular momentum, black holes
with degenerate and non-rotating horizons in five spacetime dimensions.
The fact that the horizon is non-rotating is due to supersymmetry \cite{GMTown}. Any supersymmetric solution of the supergravity equations admits
an everywhere causal killing field, which precludes an ergoregion.

The Bekenstein-Hawking entropy of the horizon has been shown to 
agree with that
of a system of D-branes and strings. 
It goes to zero as the angular momentum approaches a critical value.
In this ``under-rotating'' case (to be defined precisely in section 2) the CTC's are enclosed inside a velocity of 
light surface (VLS) which lies inside the horizon. 
Thus external observers  cannot use  the system as a ``time machine".
This is rather like the situation for rotating (and non-supersymmetric)
Kerr black holes in four spacetime dimensions.

In the over-rotating case there are CTC's outside 
what superficially appears to be a horizon. One might think that,
as in the case with an over-rotating Kerr solution in
 four spacetime dimensions, this naked time 
machine would also admit naked singularities. 
One might then be able to argue
that time travel would be excluded by the Cosmic Censorship Hypothesis,
as appears to be the case in four spacetime dimensions. 
However this is not so for our over-rotating solutions. 
In fact, as we show later, causal geodesics cannot penetrate
the region inside the  ``horizon"
and so one has a geodesically complete, simply connected, 
asymptotically flat, finite mass,
non-singular, time-orientable spacetime 
containing a naked time machine. Moreover the solution is superymmetric
and the energy momentum tensor satisfies the dominant energy condition.

This is potentially rather disturbing. However, the solution represents
a spacetime configuration which has existed for all time. 
The physically more interesting question is whether such a 
device could be constructed starting from a spacetime containing no CTC's.
According to the Chronology Protection Conjecture \cite{cpc}, this should be impossible.
Specifically it should be impossible to pass from the under-rotating 
to the over-rotating case, 
increasing the angular momentum of the spacetime and expending a 
finite amount of energy. If this were true, the situation would be similar
to that of theories admitting tachyons. One cannot speed up a sub-luminal
particle to a speed equal to, let alone greater than,
that of light, using a finite amount of energy.

\begin{figure}[h]
\begin{picture}(0,0)(0,0)
\end{picture}   
\centering\epsfig{file=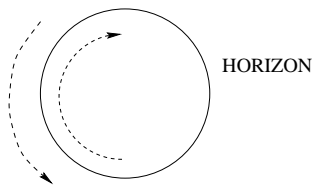,width=4cm}   
\caption{Distribution of angular momentum in the BMPV spacetime}
\label{drag}
\end{figure}

Another source of interest for our study lies in the fact that the coexistence of rotation and supersymmetry originates a rather peculiar geometry. The distribution of angular momentum must be compatible with the requirement of a non-rotating horizon, which is achieved by having a non-rigidly rotating spacetime. More specifically, it was pointed out in \cite{GMTown} that a negative fraction of the total angular momentum of the spacetime is stored behind the horizon. Hence, the fact that the horizon is static was interpreted as a cancellation of opposite dragging effects (Figure \ref{drag}).

In this paper, we will embark on an investigation of these issues. A deeper understanding of the exotic geometry and the causality violations of a BPS vacuum state being our main motivations. 
We begin our account in section 2 by discussing
the isometry group. We then discuss some global questions 
associated with the  introduction of time coordinates and time functions.
This is of particular importance because, not only does the BMPV spacetime 
have event horizons at which our usual ideas about time break down, 
but it also has closed timelike curves which 
leads to a further breakdown of familiar temporal concepts.

In section 4 we discuss the global structure and geodesic flow
of the BMPV spacetimes. Our task is then to start with the local form of the metric and obtain a maximal analytic  extension. By analytic, we mean
that the metric is invertible and its 
components  real analytic functions 
of the coordinates in  every chart of the atlas 
of charts covering 
the manifold. By maximal we mean that every causal geodesic may either
be extended for  infinite affine parameter values, in both directions, or 
encounters a singularity.  
To do this we shall start in 
a neighbourhood of spatial infinity
where the metric is asymptotically flat and extend this region
by following geodesics through the horizon, where our initial coordinates will break down.
The geodesics will either hit the curvature
singularity or pass through a Cauchy
horizon into some other asymptotically flat region. However, for an over-rotating black hole they will not be able to enter the black hole region at all.
Moreover, we discuss how the dragging effects of the spacetime allow or prevent particles in free fall to enter the Velocity of Light Surface. We show that when geodesics penetrate the VLS they can time travel, as judged by the observer at infinity, and we discuss the consistency of such time travel.

In order to follow the geodesics 
and check maximality,
it is necessary to solve for
their motion. For this we make use of the constants of the motion
arising from the large isometry group.
In addition we show, by separating variables in the 
Hamilton-Jacobi equation, that an additional 
constant of the motion arises, analogous to total angular
momentum. It's existence may be traced to the existence
of a second rank Killing tensor which we investigate in section 3.

Section 5 is devoted to the quantum mechanical problem of a spinless particle in this background. A clear distinction is then found between the over and under-rotating cases. For the former, the behaviour found for geodesics is confirmed and extended to \textit{all} frequencies: the absorption cross section, $\sigma_{abs}$, is always zero. For the latter, $\sigma_{abs}$ is found to coincide with the area of the hole, for small values of a ``frequency'' parameter, extending to the BMPV family previously known results for static and 4 dimensional Kerr black holes. These results suggest a speculation for the area and entropy of over-rotating spacetimes. For high frequencies, the WKB approximation is used to compute the absorption cross section. We find it depends sensitively on the ``azimuthal" angular quantum number of the wave. This is in contrast with the result for small frequencies, for which $\sigma_{abs}$ is dominated by the s-wave, hence insensitive to the angular momentum components.

Section 6 elaborates on the impossibility of `constructing' an over-rotating black hole, starting from an under-rotating one, under a reversible thermodynamical process. Analogies with an ordinary particle and tachyon are emphasised and a comparison with the Kerr case is drawn. We end up with a Conclusion and discussion.

\sect{The BMPV spacetime and Causality}
The BMPV spacetime is described locally by the following metric:
\bequ
ds^{2}=-\Delta^{2}\left[dt+f(d\gamma+\cos{\beta}d\alpha)\right]^{2}+\frac{dr^{2}}{\Delta^{2}}+\frac{r^{2}}{4}[d\alpha^2+d\beta^2+d\gamma^2+2\cos{\beta}d\alpha d\gamma],
\label{spmetric}
\eequ 
where 
\bequ
\barr{c}
\displaystyle{\Delta=1-\frac{\mu}{r^{2}}}, \ \ \ \ \ \displaystyle{f=\frac{\mu \omega}{2(r^{2}-\mu)}},
\earr
\eequ
and $\alpha, \beta, \gamma$ are the standard Euler angles
on $SU(2)$, with ranges $0\le \beta < \pi$, $0\le \alpha < 2\pi$ and $0\le \gamma < 4\pi$. We will also define 
\bequ
\Delta_L=(1-{ \omega^2  \mu^2 \over r^6}),
\eequ
the quantity which determines the location of the VLS. Accordingly, 
we define the positive quantities $r_H$ and $r_L$ by 
\bequ
\barr{c}
\displaystyle{\Delta=( 1-{r_H^2  \over r^2}), \ \ \ \ \ \Delta_L=(1- { r_L ^6 \over r^6} )}.
\earr
\eequ   
The complete supergravity solution also contains two one-form potentials, corresponding to a Ramond-Ramond gauge field and a winding gauge field. Both dilaton and moduli fields are constant.

According to \cite{GMTown} the ADM mass is
\ben
M={ 3 \pi \mu \over 4 G_5},
\label{ADM}
\een
and the angular momenta are
\ben
(J_L, J_R)= (0, -\pi {\omega \mu \over 2 G_5} ), 
\label{ADMJ}
\een
where $G_5$ is Newton's constant in five spacetime dimensions, which we set to one in the following.

\subsection{The isometry group}
 
The BMPV spacetime is  asymptotically
flat, stationary, with associated Killing field $\bf K$,
and invariant  under the action of $U(2)$, acting on
with three-dimensional orbits which are spacelike
near infinity.
The asymptotic flatness dictates that these orbits are 
copies of $SU(2) \equiv S^3$, rather than some quotient. 
The isometry group is therefore
${\Bbb R} \times U(2)$ with a simply transitive
four-dimensional  subgroup ${\Bbb R} \times SU(2)$ with Killing fields
$\bf K$, and ${\bf R}_1 ,{\bf R}_2,{\bf R}_3$. The last three
may be thought of as generating left translations on $SU(2)$.
The fifth Killing field, ${\bf L}_3$ may be 
identified with a  
a left-invariant vector field generating an additional right translation.
In what follows it will be important that all one parameter
subgroups of $SU(2)$ are conjugate to a circle subgroup.
Thus, for example, the orbits of ${\bf R}_1, {\bf R}_2$ and ${\bf R}_3$ 
are circles. The same remark applies  to the
the generators of right translations ${\bf L}_1, {\bf L}_2,{\bf L}_3$.
In the case of ${\bf L}_3$ the orbits are the fibres
of the Hopf fibration of $S^3$.

If  $(\sigma ^0,\sigma ^1 , \sigma ^2 , \sigma ^3)$ 
are the left-invariant one forms on ${\Bbb R} \times SU(2)$, the metric 
may be written, near infinity, as \cite{GMTown}
\ben
ds^{2}=-\Delta ^2 \left[ \sigma ^0 + {\omega \mu \over 2 \Delta r^2}
\sigma ^3\right]^2 + { dr^2 \over \Delta ^2 } + { r^2
\over 4} \left[ (\sigma ^1)^2 + (\sigma ^2)  ^2 + (\sigma ^3)^2  \right].
\een
The spacelike coordinate $r$ labels the orbits of ${\Bbb R} \times SU(2)$
and as long as $\Delta \ne 0$ the orbits of ${\Bbb R} \times SU(2)$ 
are timelike.

Because
\ben
g ({\bf K}, {\bf K}) =- \Delta ^2,
\een
the orbits of $\bf K$ are never spacelike.
To avoid a naked singularity we choose $\mu$ to be positive.
It follows 
that  $\bf K$ has degenerate Killing horizons at $r=\sqrt {\mu}$.
They are smooth null hypersurfaces.

Since
\ben
g({\bf L}_3, {\bf L}_3 ) = r^2 \Delta _L,
\een
the Hopf fibres, i.e. the orbits of ${\bf L}_3$, are timelike  
if $\Delta_L <0$, 
which is thus a region of closed timelike curves. 
In this region, we shall say {\it inside the time machine}, 
the induced metric on the orbits of $SU(2)$ 
is Lorentzian. In fact the metric inside 
the time machine is similar to the 
outer reaches of the Lorentzian Taub-NUT metric. 

The local
metric form (\ref{spmetric}) is valid everywhere
inside and outside the horizon, including 
at points where $\Delta_L=0$. Thus, 
the boundary 
of the time machine, which we call the \textit{Velocity of Light Surface} (``VLS'') 
is a smooth non-singular time-like hypersurface on which 
the Killing field ${\bf L}_3$ becomes null.

Because 
\ben
g({\bf K}, {\bf L}_3)= -\Delta { \omega \mu \over r^2}
\een
inside the time machine, the timelike Killing field
${\bf L}_3$ is  future pointing
or past pointing depending on the sign of $ \mu \omega  \Delta$.
If one takes the sign of $\mu \omega$ to be positive,
these two cases correspond to whether the time machine lies inside the
horizon (i.e. $\omega < \mu ^{1\over 2}$ or $r_L < r_H$ ), which we shall 
refer to as the {\it under-rotating } case,
or the horizon lies inside the time machine 
(i.e. $\omega > \mu ^{1 \over 2}$ or $r_H< r_L$ ), which we shall
refer to as the {\it over-rotating } case.

The metric admits additional
discrete symmetries of which the most important for us
is that  corresponding to the simultaneous reversal of
${\bf K}$ and  ${\bf L}_3$, i.e., to the simultaneous reversal
of $\sigma ^0$ and $ \sigma ^3$. Thus, the 
simultaneous reversal of time and sense of rotation
of the system leads to an indistinguishable configuration. This invariance corresponds to PT symmetry.

\subsection{Time orientation and time functions} 

Intuitively a time orientation on a spacetime $M$ is a continuous 
assignment  of a {\sl future} light cone at each point of spacetime.
Given such an assignment, one may tell whether  an (oriented)
causal curve is moving to the future or to the past.
If future directed casual curves  are identified with 
the worldlines of classical particles then past directed 
causal  curves may, \`a la St\"uckelberg and Feynman,
be identified with the world lines of
classical  anti-particles. If spacetime lacked a time orientation it would
be impossible to distinguish classical particles from classical
anti-particles. The lack of a time orientation would
preclude a conventional
  quantum field theory in the background spacetime
since it leads to real quantum mechanics \cite{Gib}. 
It is reassuring, therefore,
that the BMPV spacetime admits a global
time orientation.

Spacetime may only fail to 
be time orientable if it is not simply connected.
Moreover one may always pass to a double cover of
a non-time-orientable spacetime to obtain a spacetime 
which is time orientable. However in a concrete case
it may not be easy
to determine the local direction of time. This is specially
true if the spacetime does not admit a time function.
The simplest way to provide a time-orientation is
to give an everywhere non-vanishing causal vector field
$\bf V$ which is deemed to point to the future.
Obviously $-{\bf V}$ is deemed to point to the past. 
Thus, and despite its closed timelike curves,
the 
BMPV spacetime is time orientable.
One may provide it with
a time orientation by choosing for $\bf V$ the Killing field $\bf K$, 
since
\ben
g({\bf K}, {\bf K} )= -(1- {\mu \over r^2 } ) ^2, 
\een
$\bf K$ is everywhere causal and never vanishes. Note that
for an ordinary, non-extreme, black hole we could not 
use the time-translation Killing field to provide 
a time orientation because it is not everywhere causal
and moreover it vanishes on the Boyer-Kruskal axis.

The fact that one has a time orientation is no guarantee that
one has a global time function. The latter is usually defined as a 
continuous function
$t: M \rightarrow {\Bbb R}$ with $dt \ne 0$
whose level sets are spacelike hypersurfaces and which
each causal curve crosses once and only once.
Clearly a spacetime, such as the BMPV spacetime,
with closed timelike curves cannot 
admit a global time function.
The best one can do is to construct a local time function.

\subsection {Local Time functions adapted to a Killing field}

Given a causal Killing vector
$\bf K$, one might try to find  a time function $t$ by solving the equation
\ben
{\bf K} t=1.\label{time}
\een
The equation (\ref{time}) is a first order partial differential 
equation which one may solve as follows
in a small neighbourhood $U \subset M$
in which $\bf K$ is timelike.
One picks as
an initial level set of the function $t$  
an arbitrary  spacelike hypersurface in $U$, call it $s$, transverse to $\bf K$, and assigns it the value
$t=0$. The function $t$ is then extended off $s$ by
solving equation (\ref{time}) along the integral curves of $\bf K$.
If  $x^i$ are local coordinates on $s$, they may be extended
to the neighbourhood $U$  by the stipulation that they be 
co-moving, that is, constant
along the integral curves of $\bf K$:
\ben
{\bf K} x^i =0
.\een
In the neighbourhood
$U$ the metric then may be given the form
\ben
ds^2 = -V^2 (dt + A_i dx^i)^2 + \gamma_{ij}dx^i dx^j, \label{metric}
\een
where the quantities $ V$, $A_i$ and $\gamma_{ij}$ depend 
only on the coordinates $x^i$. Changing the initial hypersurface $s$
corresponds to the ``gauge transformation"
\ben
t \rightarrow t+ f(x^i),
\een
under which
\ben
A_i \rightarrow A_i - \partial_i f,
\een
but the function $V$ and the metric $\gamma_{ij}$ remain unchanged. 
They are invariantly defined as the length of the Killing field $\bf K$
and the projection of the spacetime   orthogonal to $\bf K$
respectively. The metric induced on $s$, $g_{ij}=\gamma_{ij}- V^2 A_i A_j$, is not gauge-invariant since it depends 
upon the choice of $s$. A time function
corresponds to an initial hypersurface $s$ such that the metric $g_{ij}$
induced on it
is positive definite. 
If this is not so we merely have a {\it time coordinate}. Some of the apparent puzzles and paradoxes connected with time travel may be resolved if one bears in mind the distinction between a time coordinate and a global time function. Whenever CTC's are present, only the former can exist. But the ability to choose some arbitrary initial data on a $t=constant$ hypersurface relies on it being a Cauchy surface, which is only the case if $t$ is a global time function. This remark will be illustrated later.

Note that if 
\ben
F_{ij}= \partial _i A_j- \partial _j A_i =0,
\een
then one may pick an $f$ such that $A_i=0$. This is the static case
for which the one form $g({\bf K}, )=  -V^2 d t$. 
The hypersurface  $s$ is picked out, up to a 
time translation,  by the condition that it ought to be
orthogonal to the integral curves of $\bf K$.

\subsection{Spacetime as a real line bundle and the Sagnac connection}

There are several reasons why the procedure above
might fail globally. 
For example, the Killing field may vanish somewhere in the spacetime 
$M$. This does not happen for the BMPV spacetime so we 
ignore this possibility. It then follows that we may 
regard the spacetime $M$
as a principal  $\Bbb R$ bundle over
a base space $B$ which may be thought of as the space of orbits
of the time translation group. The projection map $\pi: M \rightarrow B$ is
the assignment to each point in spacetime of the unique integral curve of
$\bf K$ passing through that point.  
This bundle might be called the gravitomagnetic or Coriolis
bundle, with structural group given by time translations.

If the orbits of $\bf K$ are everywhere timelike, the metric $g$
on the bundle
$M$ determines a connection, let us call it the Sagnac connection \cite{sagnac},
such that a horizontal curve is
one which is
orthogonal to the integral curves of $\bf K$.

In this language the construction of a time function
amounts to picking a local section $s$ of the bundle, which is moreover
required to be spacelike. 
The pull back of the curvature of the Sagnac connection to $B$
is just the two-form $F_{ij}$.
If the time translation
group were the  circle $S^1$ rather than the line $\Bbb R$, it is possible
that no global section of the bundle
exists. However this cannot happen if the 
group is $\Bbb R $.

What can happen, however, is that the Killing field $\bf K$ can fail
to be everywhere timelike. Thus the metric form (\ref{spmetric})
may break down, and the Sagnac connection become ill-defined. 
In the case of the BMPV black hole this happens 
where $\Delta=0$. Thus we cannot extend the time coordinate
$t$ beyond the horizon at $r=\sqrt \mu$. Inspection of the metric form
(\ref{spmetric}) shows that, apart from the usual 
angular coordinate singularities of Euler angles,
 $(t,r, \alpha, \beta, \gamma)$ 
are good coordinates throughout the exterior of the hole, $r>\sqrt\mu$.
One may also use the same form but where $t$ is not the same coordinate
throughout the region in which $0<r<\sqrt \mu$. 
We shall call these two charts the 
exterior patch $U_{\rm ext}$ and interior patch $U_{\rm int}$
respectively. Neither covers  the horizon. 
Thus they certainly do not constitute a complete atlas.

For the BMPV black hole the metric $\gamma_{ij}$
orthogonal to
the orbits of $\bf K$ is given by
\ben
 \gamma_{ij}dx^i dx^j ={ dr^2 \over \Delta ^2 } + { r^2
\over 4} ( (\sigma ^1)^2 + (\sigma ^2)  ^2 + (\sigma ^3)^2  ),
\een
while if $\sigma ^0=dt$ then
 $g_{ij}$ is
\ben
g_{ij}dx^i dx^j= \frac{dr^2}{\Delta^2}+{ r^2 \over 4} (  (\sigma ^1)^2 + (\sigma ^2)  ^2 + \Delta _L
(\sigma ^3)^2  ).
\een
It is clear, therefore, that the time coordinate $t$ ceases to be a time
function inside the time machine, since the metric $g_{ij}$
becomes timelike there.

In the over-rotating case we can use $t$ as a 
time coordinate outside the horizon but we shall find that
a future directed timelike curve, indeed a  geodesic, 
can move backwards in that time coordinate.

The Sagnac connection of the BMPV metric is 
\ben
A_idx^i  =  {\omega \mu \over 2 r^2 \Delta} \sigma ^3,
\een
which is perfectly well defined on the timelike hypersurface which bounds
the time machine. If one computes the Sagnac curvature
one finds that it is anti-self dual with respect to the four-dimensional
conformally flat metric $\gamma_{ij}$. 
This becomes clearer if one introduces isotropic coordinates
${\bf x}$, with $|{\bf x}|= \sqrt {r^2 -\mu}$.
If $H$ is the harmonic function on 
four-dimensional Euclidean space ${\Bbb E}^4$ 
given by
\bequ
H= 1 + { \mu \over {\bf x}^2 },
\eequ 
then the BMPV metric may be written as

\ben
ds^2 =H^{-2} (dt + A_i dx^i)^2 + H d {\bf x}^2, \label{harmonic}
\een
where
\ben
A_i= -\frac{\omega}{2} I^3_{ji} \partial_j H,
\een
and where $ I^3_{ij}$ is a constant self-dual form
on ${\Bbb E}^4$. The Sagnac curvature is then given by
\ben
F_{ij}=\frac{2\mu \omega}{\bf{x}^{4}}\bar{I}^{3}_{ij},\een
where $ \bar{I}^{3}_{ij}$ is an anti-self-dual tensor.

The isotropic coordinates cover the exterior
region in both the under and the over rotating case.
One obtains a solution representing $k$ black holes
by taking the harmonic function
$H$ to have $k$ poles \cite{gauta}.

\subsection{Axi-symmetric Stationary Time machines}

In order to facilitate comparisons
with some well known and more familiar spacetimes with closed
timelike curves it will prove useful to develop
a general formalism valid for stationary spacetimes which 
admit, in addition, a circle action. The  Killing field 
generating the circle action will be called $\bf m$.
It is well defined by demanding the integral curves to be closed circles. 
The circle group is given by the coordinate $\phi$, with range 
$0\le \phi \le 2\pi$.  
We will, in addition, assume that the metric is invariant under 
the simultaneous
reversal of $\bf K$ and $\bf m$. 
In other words, that the action of ${\Bbb R} \times S^1$
is orthogonally transitive. In the case of the BMPV black hole, 
$\phi=\gamma$ and ${\bf m}={\bf L}_3$.

If $\phi$ is a conventional angle,
this symmetry corresponds to $PT$ symmetry. However 
the formalism works perfectly well
if $\phi$
parameterises an internal or higher dimensional
angle \`a la Kaluza-Klein. One is then assuming $CT$ symmetry.
In this interpretation, angular momentum corresponds to
electric charge and angular velocity corresponds to 
electrostatic potential.
 
Under these assumptions, the metric may be written as
\ben
ds^2= -Vdt^2 + 2 W dt d \phi + X d \phi ^2 + g_{ab} dx^a dx^b,
\een
where $a=1,\dots,n-2$ 
are coordinates on the orbit space $N=M/ ({\Bbb R} \times S^1)$,
and the functions $V,X,W$ depend 
on $x^a$. Ergospheres are given by $V=0$ and the 
velocity of light surface is given by $X=0$. The metric on the $t-\phi$ plane has determinant $-\Sigma^{2}$, with
\ben
\Sigma ^2 = VX + W^2.
\een
Hence, at points where  $\bf K$ and $\bf m$ are non-zero and 
linearly independent,
the 2-surfaces of transitivity of ${\Bbb R} \times S^1$
are timelike as long as $\Sigma^{2} >0$. In that case any  ergospheres
and velocity of light surfaces are timelike.
The 2-surfaces of transitivity  can become
light-like, corresponding to a Killing horizon of some linear combination 
${\bf l}={\bf K} + \Omega_H {\bf m}$ of the two Killing vectors. This corresponds to a Killing horizon
with null generator $\bf l$.
A necessary condition is that   $\Sigma^2=0$.
The quantity $\Omega_H$ is the angular velocity of the horizon. 
In the case of the 
BMPV black hole one has $\Sigma^{2}=\left(r\Delta /2\right)^{2}$, and since the horizon is at  $r=r_H$, we have $\Omega_H=0$.

One may express the metric as
\ben
ds^2= -V(dt-{W \over V}d\phi)^2 + { \Sigma ^2 \over V} d\phi ^2 + g_{ab} dx^a dx ^b.
\een
Thus the Sagnac connection is given by
\ben
A= -{W \over V}d \phi.
\een

The energy $E$ and angular momentum $j$ of a particle 
whose worldline has affine parameter $\lambda$
are given by
\ben
E=V{ dt \over d \lambda} - W { d\phi \over d \lambda}
\een
and
\ben
j=X{ d\phi \over d \lambda} + W { dt \over d \lambda},
\een
which can be rewritten as 
\ben
{dt \over d \lambda}= { X\over \Sigma ^2} (E+{W \over X} j) 
\label{dtgen}
\een
and
\ben
{d\phi \over d \lambda}= { V\over \Sigma ^2} (j-{W \over V} E).
\een
If $\Sigma ^2 \ge 0$,
as is the case for the BMPV black hole, and when $X$ is positive, $E$ must be positive for 
a particle moving forwards in time and negative for  a
particle moving backwards in time. The latter can be though of as an antiparticle. However, as we shall see,
a particle with positive energy $E$ will not necessarily
move forwards with respect to the coordinate $t$. 
Inside the time machine, where $X$ is negative, it may be possible for 
the particle to move backwards
with respect to the coordinate $t$. Clearly, the energy term is then negative, but the phenomenon is actually a consequence of a detailed balance between the two terms in (\ref{dtgen}).

It is useful to define a quantity
\ben
V^0= -{W \over X}j.
\een
It tells us whether particles move forward or backward
with respect to the coordinate $t$.
If we suppose that $\Sigma ^2 >0$, then a particle with positive
energy $E$ will move forward with respect to the coordinate
$t$ as long as 
\ben
E > V^0
\label{condc}
\een
outside a  time machine, (i.e. if $X>0$). Of course, this must be the case, since the light cone structure of the spacetime does not allow classical causal particles to move backwards in time outside the time machine. Thus, (\ref{condc}) must follow from causality. This is illustrated in section 4.5. Inside a time machine
(i.e. if $X<0$), to move forward with respect to $t$ one must have
\ben
E < V^0.
\een
The metric dependence of the quantity $V^0$ also governs the dragging of inertial frames.
For example, a particle  with zero angular momentum will be 
dragged around in $\phi$ as
\ben
\frac{d \phi}{ d \lambda}= V^0 {dt \over d \lambda}.
\een
 
A simple calculation shows that
\ben
g_{ab} { dx ^a \over d \lambda} { dx ^b \over d \lambda} =
{ X \over \Sigma ^2} (E-V^+) (E-V^-)
\een
where
\ben
V^{\pm}= V^0 \mp \sqrt{ \left(m^2 +\frac{j^2}{X}\right)\frac{\Sigma^{2}}{X}}. 
\een
The quantities $V^{\pm}$ serve as effective potentials.
Consider the space ${\Bbb R} \times N$ with coordinates ($E,x^a$),
where $N$ is the orbit space =$M/({\Bbb R} \times S^1)$.
Outside the time machine the allowed values of energy lie 
above or below the surfaces $E=V^{\pm}$ while
inside a time machine the allowed values of $E$ lie between the 
surfaces $E=V^{\pm}$.  The surface $E=V^0$ lies between the 
the surfaces $E=V^{\pm}$. Inside the time machine, 
it  bounds the region in which the particle
moves backwards with respect to the coordinate $t$. 

The same quantity governs solutions $\Phi$
of the  Klein-Gordon equation. If $\Phi = \chi(x^a) e^{-i(Et-j \phi)}$,
one finds that
\ben
-{1 \over \Sigma  } \nabla _a ( \Sigma  g^{ab} \nabla _a \chi)
= { X \over \Sigma ^2} ( E-V^+)(E-V^-) \chi.
\een

In the case of the BMPV black hole, where there exists
an additional set of constants
of the motion, one may refine the idea of an effective potential.
By using the constants one gets a modified
 effective potential
which depends only on the radial variable.
In effect the extra constants are included in an effective mass term.
The results about the allowed regions and travel
backwards in time go through in a similar way.

\subsection{Ingoing null coordinates}

In order to pass through Killing horizons one may
introduce an ingoing or advanced null time coordinate  $v$ satisfying the conditions
that
\ben
{\bf K}v=1,
\een
and the surfaces of constant $v$ are null hypersurfaces.
It is convenient to change the Euler coordinate $\gamma$ at the same time.
Thus one defines
\ben
dv=dt + f(r)dr,
\label{vtime}
\een
\ben
d\gamma_{\rm in} = d \gamma + g(r)dr,
\label{gamin}
\een
and hence 
\ben
{\sigma ^3}_{\rm in}  =\sigma +g(r)dr, 
\een
where the functions $f(r)$ and $g(r)$ are chosen to render 
zero the metric
coefficients of $dr^2$ and ${\sigma ^3}_{\rm in}  dr$.
It follows that not only are the surfaces $v={\rm const}$ 
ingoing null hypersurfaces
but that the null generators (which are in fact ingoing null geodesics)
are given by  $\gamma_{\rm in} = {\rm constant}$, $\beta = 
{\rm constant}$, $\alpha = {\rm constant}$. One must choose
\ben
f=\frac{\Delta_{L}^{\frac{1}{2}}}{\Delta^{2}}
\een
and
\ben
g=\frac{2\mu \omega}{\Delta r^{4}}\Delta_{L}^{-\frac{1}{2}}.
\een

The metric becomes 
\ben
ds^{2}=-\Delta dv^2 -dv\left[2\Delta_{L}^{-\frac{1}{2}}dr+\frac{\mu \omega \Delta}{r^{2}}{\sigma ^3}_{\rm in}\right]+ { r^2 \over 4} \left[ (\sigma ^1 )^2  +(\sigma ^2)^2 + \Delta _L({\sigma ^3}_{\rm in} )^2      \right] ,
\een
where we have used the identity that
\ben
({\sigma ^1 }^\prime )^2  +({\sigma ^2} ^\prime
)^2=(\sigma ^1 )^2  +(\sigma ^2)^2.
\een

By time  reversal invariance there 
is an analogous  set of outgoing or advanced null coordinates
$u, \gamma _{out} $, obtained by reversing the signs of $f$ and $g$.

The ingoing or outgoing
null coordinate systems break down as one 
reaches the boundary of the time machine. 
This is because the null generators
of the null hypersurfaces $v={\rm constant}$
are ingoing null geodesics with zero angular momentum. 
As as we shall see later, 
ingoing null 
geodesics with zero total angular momentum cannot enter the time machine.
However if the time machine lies inside the horizon, then
ingoing and outgoing null coordinates  remain non-singular as one crosses
the horizon. In what follows we shall call these two types of patches 
$U_{\rm in}$ and $U_{\rm out}$.

This information is almost sufficient to construct the maximal extension
in the under-rotating case.
We start in the exterior with an ingoing patch which takes us through
the horizon into the region between the horizon and the time machine.
We now use an outgoing patch to take us outside into 
another exterior patch.

\sect{St\"ackel-Killing tensor}
It is well known that spacetime isometries give rise to constants of motion along geodesics. However, the converse statement is not true in 
general: not \textit{all} conserved quantities along geodesics arise from an
 isometry of the manifold and associated Killing vector field. Such integrals of motion are related to ``hidden'' symmetries of the manifold, which manifest themselves as tensors of rank $n>1$, satisfying a generalized Killing condition, namely $D_{(\mu}K_{\nu_{1}..\nu_{n})}=0$. They are usually referred to as \textit{ St\"ackel-Killing tensors}. One way to think of such "hidden" symmetries is as homogeneous functions in momentum  of degree $n$, $\mathcal{K}^{(n)}(p)$, which are defined in phase space, 
and which commute with the Hamiltonian in the sense of Poisson brackets.
In the cotangent bundle,  all $\mathcal{K}^{(n)}(p)$ are on equal footing, including the $n=1$ case. However, projection to base space introduces an asymmetry, since only linear functions in momentum will survive, and hence Killing
 vectors and tensors seem to play different roles in configuration space.

We now compute the most general St\"ackel-Killing tensor for spacetime (\ref{spmetric}),  by finding the most general quadratic form in momentum that commutes with the Hamiltonian for a scalar, uncharged particle. It is convenient to use orthonormal frames, with the obvious choice being:
\bequ
\barr{c}
\displaystyle{{\bf e}^{t}=\Delta(dt+f\sigma^{3}), \ \ \  \ \ \ {\bf e}^{r}=\frac{1}{\Delta}dr, \ \ \  \ \ \ {\bf e}^{i}=\frac{r}{2}\sigma^{i}},
\label{frames}
\earr
\eequ
and dual basis:
\bequ
\barr{c}
\displaystyle{{\bf e}_{t}=\frac{1}{\Delta}\frac{\partial}{\partial t}, \ \ \  \ \ \ {\bf
e} _{r}=\Delta\frac{\partial}{\partial r}, \ \ \  \ \ \ {\bf e}_{i}=\frac{2}{r}({\bf L}_{i}-\delta_{3i}f\frac{\partial}{\partial t})},
\label{framesinv}
\earr
\eequ
where the ${\bf L}_{i}$ are the left invariant vector fields on $SU(2)$. 
Recall that the ${\bf L}
_{i}$ generate the right action and therefore only one of them, ${\bf
L} _{3}$, is a Killing vector field for (\ref{spmetric}).

The Hamiltonian can then be written as:
\bequ
\barr{l}
\displaystyle{H=\frac{1}{2}g^{\mu\nu}p_{\mu}p_{\nu}= \frac{1}{2}\eta^{ab}e_{a}^{\mu}e_{b}^{\nu}p_{\mu}p_{\nu}}
\\\\ \ \ \ \ \displaystyle{=\frac{1}{2}(-\frac{1}{\Delta^{2}}+\frac{4f^{2}}{r^{2}})E^{2}-\frac{4f}{r^{2}}E\mathcal{M}_{3}+\frac{1}{2}\Delta^{2}p_{r}^{2}+\frac{2}{r^{2}}(\mathcal{M}_{1}^{2}+\mathcal{M}_{2}^{2}+\mathcal{M}_{3}^{2})}.
\earr
\eequ
The moment maps are $E=K^{\mu}p_{\mu},\mathcal{M}_{i}=L_{i}^{\mu}p_{\mu}$ where $K^{\mu}$ are the components of the timelike Killing vector field. Expressing a general function $\mathcal{K}^{(2)}$ in phase space as
\bequ
\barr{l}
\displaystyle{\mathcal{K}=\frac{1}{2}K^{\mu\nu}p_{\mu}p_{\nu}= \frac{1}{2}K^{tt}E^{2}+\frac{1}{2}K^{rr}p_{r}^{2}+\frac{1}{2} K^{ij}\mathcal{M}_{i}\mathcal{M}_{j}+ K^{tr}Ep_{r} +}
 \\\\ \ \ \ \ \ \ \ \ \ \ \ \ \ \ \ \ \ \ \ \ \ \ \  \displaystyle{+K^{ti}E\mathcal{M}_{i}+K^{ri} p_{r} \mathcal{M}_{i}},
\earr
\eequ
the computation of $\{H,\mathcal{K}\}_{PB}=0$ becomes straightforward. One only has to note that the only non-trivial brackets come from
\bequ
\barr{c}
\{\mathcal{M}_{i},\mathcal{M}_{j}\}_{PB}=-\epsilon_{ij}^{ \ \ k}\mathcal{M}_{k} \ \ \ , \ \ \ \{P_{r},r\}_{PB}=1.
\earr
\eequ
Notice that the components of $K$ are effectively written in the basis $(\partial /\partial t, \partial / \partial r, {\bf L}_{i})$, and we assume only $r$ dependence, since the inverse metric in such basis depends only on the radial coordinate. Under such conditions the most general $\mathcal{K}$ commuting with $H$ is found to have the form
\bequ
\mathcal{K}=Ag^{\mu\nu}p_{\mu}p_{\nu}+Bp_{t}^{2}+Cp_{t}\mathcal{M}_{3}+D(\mathcal{M}_{1}^{2}+\mathcal{M}_{2}^{2})+E\mathcal{M}_{3}^{2},
\eequ
where $A,B,C,D,E$ are arbitrary constants.

The result is both simple and enlightening. As should be expected, the metric and the Killing vector fields arise. But the requirement that the terms involving ${\bf L}_{1}$ and ${\bf L}_{2}$ should have the same coefficient is a non-trivial restriction, which would not exist for the spherically symmetric case. It can be understood by noticing that only such a sum can be expressed as a sum of products of Killing vectors, namely:
\bequ
{\bf L}_{1}^2+{\bf L}_{2}^2= {\bf R}_{1}^2+{\bf R}_{2}^2+{\bf R}_{3}^2-{\bf L}_{3}^2.
\eequ
Hence, we learn that any second order Killing tensor we might built is, unlike in the four dimensional Kerr-Newman case, reducible. Furthermore, it must include the left invariant vector fields in the combination $\mathcal{M}_{1}^{2}+\mathcal{M}_{2}^{2}$, or, equivalently, the right invariant vector fields in the combination $\mathcal{N}_{1}^{2}+ \mathcal{N}_{2}^{2}+\mathcal{N}_{3}^{2}$, where $\mathcal{N}_{i}={\bf R}_i^{\mu}p_{\mu}$. Since we want to give it an interpretation of generalized total angular momentum, it seems natural to complete the Killing tensor and define it as:
\bequ
\mathcal{K}=\mathcal{M}_{1}^{2}+ \mathcal{M}_{2}^{2}+\mathcal{M}_{3}^{2}.
\label{stack}
\eequ
Clearly, it is the Casimir invariant of any of the $SU(2)$ subgroups of the $SO(4)$ rotation group in four spatial dimensions. These two Casimirs coincide for the scalar representation. Therefore, the reduction of symmetry from $SO(4)$ to $U(2)$ due to rotation, does not affect the number of constants of motion (for scalar particles) coming from angular isometries. There will still be three quantum numbers, associated with the maximal set of commuting operators. This fact will prove useful in separating variables for the geodesics and scalar field equations, which can be achieved in a way similar to the spherically symmetric case.

One can then convert (\ref{stack}) back to coordinate basis, finding that its covariant components are:
\bequ
K_{\mu\nu}dx^{\mu}dx^{\nu}=\left[\Delta^{2}f dt-\frac{r^{2}}{4}\Delta_{L}\sigma^{3}\right]^{2}+\frac{r^{4}}{16}\left[(\sigma^{1})^{2}+(\sigma^{2})^{2}\right].
\eequ

\sect{Geodesics and global structure}
The Hamilton-Jacobi equation for our spacetime is
\bequ
-\frac{1}{\Delta^{2}}(\frac{\partial S}{\partial t})^{2}+\Delta^{2}(\frac{\partial S}{\partial r})^{2}+\frac{4}{r^{2}}(({\bf L}_{1}S)^{2}+({\bf
L}_{1}S)^{2})+\frac{4}{r^{2}}({\bf L}_{3}S-f\frac{\partial S}{\partial t})^{2}=-m^{2},
\label{hj}
\eequ
where $S(x^{\mu})$ is the action function. The spacetime symmetries associated with the Killing vector fields ${\bf K}$, ${\bf L}_{3}$ and ${\bf R}_{3}$ suggest the ansatz
\bequ
S=-Et+ H(\alpha, \beta, \gamma) + W(r),
\eequ
where $H(\alpha, \beta, \gamma)= j_L\alpha+j_R\gamma+\chi(\beta)$, and $E$, $j_L$ and $j_R$ are three integrals of motion. Using further the ``hidden'' symmetry encapsulated in the St\"ackel-Killing tensor, we introduce another conserved quantity, $j^{2}$, with the dimensions of angular momentum squared, such that
\bequ
({\bf L}_{1}H)^{2}+ ({\bf L}_{2}H)^{2}+ ({\bf L}_{3}H)^{2}=j^{2}. \label{Casimir}
\label{jota2}
\eequ
This relation allows us to determine the variation of $\chi$:
\bequ
\frac{\partial \chi}{\partial \beta}= \pm \sqrt{j^{2}-\frac{1}{sin^{2}\beta}[j_R^{2}+j_L^{2}-2j_Lj_Rcos\beta]} \label{chi}.
\eequ
The equations of motion for a test particle follow from the fact that its momentum is the gradient of the action function:
\bequ
\barr{l}
\displaystyle{\frac{dr}{d\lambda}=\pm\left[(E^{2}-m^{2})+\frac{2}{r^{2}}(\mu m^{2}-2j^{2})+\frac{\mu}{r^{4}}(8j^{2}-4\omega Ej_R-m^2\mu)-\right.}
\\\\ \ \ \ \ \ \ \ \displaystyle{\left. -\frac{\mu^{2}}{r^{6}}(4j^{2}-4\omega Ej_R+\omega^{2}E^{2})\right]^{\frac{1}{2}}.}
\label{eqmr}
\earr
\eequ
Notice in the second term the characteristic signature of 5 dimensions, where the Newtonian potential has the same $r$ dependence as the repulsive centrifugal term, which is the origin for the non existence of bound states in the 5D Tangherlini black hole \cite{tang}.

The remaining equations of motion are:
\bequ
\frac{dt}{d\lambda}=\frac{E}{\Delta^{2}}\Delta_{L}-\frac{2\mu \omega j_R}{\Delta r^{4}},
\label{dtdl}
\eequ
\bequ
\frac{d\beta}{d\lambda}=\frac{4}{r^{2}}\frac{\partial \chi}{\partial\beta},
\label{beta}\eequ
\bequ
\frac{d\alpha}{d\lambda}=\frac{4}{r^{2}sin^{2}\beta}(j_L-j_Rcos\beta),
\label{alpha}
\eequ
\bequ
\frac{d\gamma}{d\lambda}=\frac{4}{r^{2}}\left[\frac{j_R-j_Lcos\beta}{sin^{2}\beta}+\frac{E\mu \omega}{2r^2\Delta}\right]. \label{gamma}
\eequ
These equations exhibit the behaviour one would expect for a rotating background. Putting the $j$, $j_R$ and $j_L$ to zero, there is no motion in the angles $\alpha$ and $\beta$, but there is angular motion in $\gamma$, corresponding to the dragging of frames caused by rotation. In other words, rotation is not the same as possessing angular momentum, when one deals with a rotating background. Moreover, the dragging changes sign from inside to outside the horizon, which confirms the picture presented in Figure \ref{drag}. From the $t$ equation, we recover, for $\omega=0$, the usual interpretation that particles ($E>0$) propagate forwards in time and antiparticles ($E<0$) backwards. For non-zero rotation, however, the energy term reverses the usual sign, as we get inside the VLS, leading to time travel; furthermore, a spin-orbit coupling term arises, which, as we show below, ``conspires'' to minimise the time travel effect along a geodesic. Notice the singularities on the horizon for the $t$ and $\gamma$ equations. In what follows these will be shown to be just a failure of coordinates.

\subsection{The angular motion and the Clifford tori}
Before discussing the motion in $r$ and $t$ we will say a few words about the motion
in the Euler angles. We begin by recalling that $S^3 \equiv SU(2)$
is foliated by  tori, called Clifford Tori, which are 
given by $\beta= {\rm constant}$. More specifically, the metric on $S^3$ can be written, by defining $u=(\alpha+\gamma)/2$ and $v=(\alpha-\gamma)/2$, as 
\bequ
ds^2=\frac{1}{4}\left[d\beta^2+4\cos^2\left(\frac{\beta}{2}\right) du^2+4\sin^2\left(\frac{\beta}{2}\right) dv^2\right].
\eequ
There are two singular leaves 
corresponding to $\beta=0$ and $\beta={\pi}$. They are circles. 
If one uses the standard round metric on $S^3$, 
each torus is intrinsically flat and in general rectangular.
But for $\beta=\pi/ 2$ the torus is a square torus, and moreover it is
a minimal two-surface with respect to the round metric on $S^3$.
If $\beta \ne \pi / 2$ the tori are not minimal
submanifolds except for the cases $\beta=0$ and $\beta=\pi$ which are 
circular geodesics

\begin{figure}
\begin{picture}(0,0)(0,0)
\end{picture}   
\centering\epsfig{file=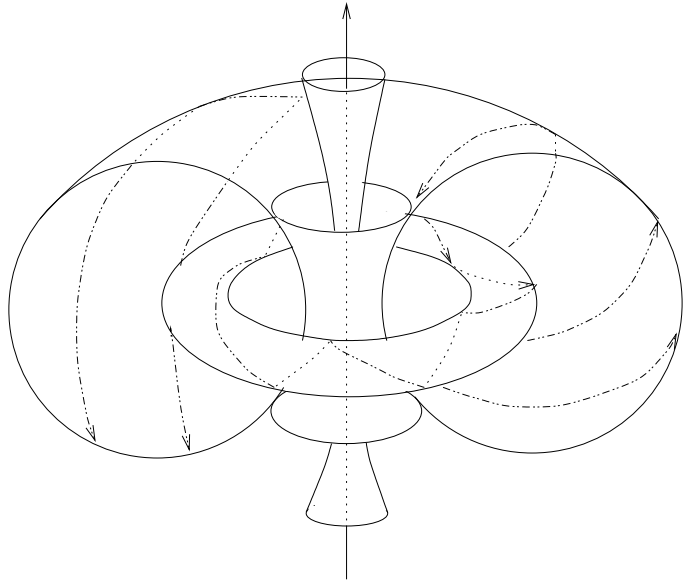,width=9cm}   
\caption{Clifford Tori; the lines correspond to the orbits of $\frac{\partial}{\partial \gamma}$.}
\label{torus}
\end{figure}

Each 
Clifford torus is fibred by two sets of circles $\alpha={\rm constant}$
and $\gamma={\rm constant}$ which  spiral around the tori
in opposite senses. They are the orbits of the
right or left actions of $U(1) \subset SU(2)$ respectively.
In other words they are the orbits of the self-dual or anti-self-dual
Hopf fibrations respectively. 
The picture becomes easy to visualize if one performs a 
stereographic projection from $S^3$ to ${\Bbb E}^3$. The tori
are then the same as when using standard toroidal coordinates (Figure \ref{torus}).

\subsection{Equatorial motions and consistent back reaction}

It follows from (\ref{Casimir}) that
\ben
j^2 \ge j_R^2. \label{jbound}
\een
Equality in (\ref{jbound}) implies from (\ref{jota2}) that $H$ depends only on $\gamma$ and hence $j_L=0$. It then follows from 
(\ref{beta}) that the motion is confined to the square torus
at $\beta=\pi/ 2$ and moreover from (\ref{alpha}) we deduce
that $\alpha ={\rm constant}$. Thus the motion traces out a Hopf fibre on the
square Clifford torus on $S^3$. Conversely every motion
with $j_L=0$ has this character. These motions, which we will refer
to in the sequel as {\it equatorial}, are particularly important.
The reason is that the BMPV black hole or time machine
has momenta given by $(J_L, J_R)= (0, -\pi \omega \mu^{\frac{1}{2}}/( 2 G_5) )$. It follows that
if it absorbs a particle with $j_L \ne 0$ it cannot remain in the
BMPV family of solutions. In fact it must become a non-BPS state.
On the other hand if the absorbed particle has $j_L=0$ then
it can remain within the  BMPV family, 
but with new left angular momentum
and mass given by $M+E$ and $J_R + j_R$ respectively. 
In this way we can gain some insight into the back-reaction problem.

\subsection{Effective Potentials}
\begin{figure}[p!]
\centering\epsfig{file=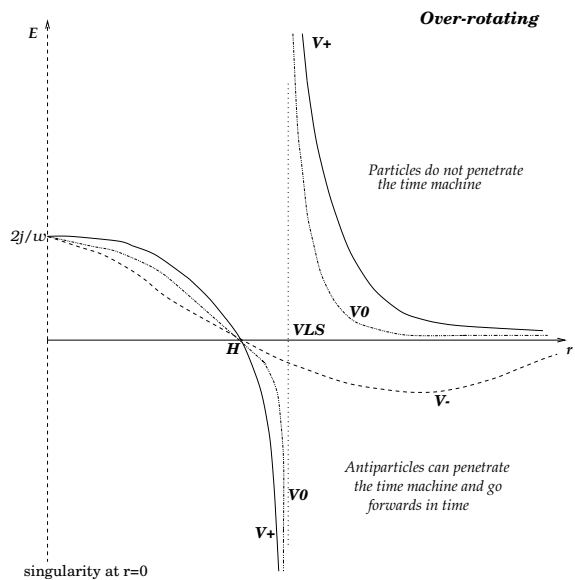,width=7.5cm}   
\caption{Effective potentials for a massless particle.}
\label{pot1}
\end{figure}

\begin{figure}[p!]
\centering\epsfig{file=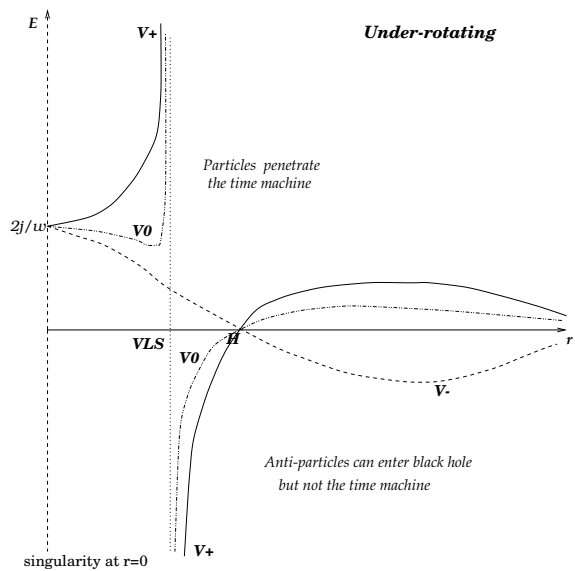,width=7.5cm}   
\caption{Effective potentials for a massless particle. Notice the special orbit with $E=\frac{2j}{\omega}$ that may reach the singularity.}
\label{pot2}
\end{figure}

\begin{figure}[p!] 
\centering\epsfig{file=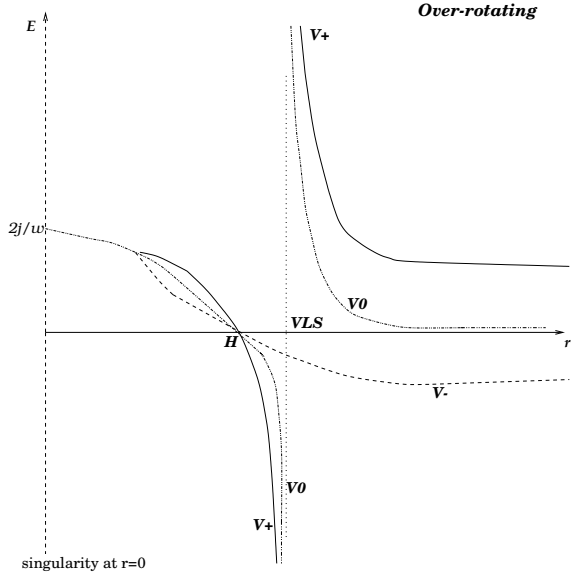,width=7.5cm}   
\caption{Effective potentials for a massive particle. As the mass increases, the ``tongue'' representing the allowed region inside the horizon gets smaller, until eventually vanishing. Furthermore, there is now a mass gap as $r\rightarrow \infty$.}
\label{pot3}
\end{figure}

\begin{figure}[p!] 
\centering\epsfig{file=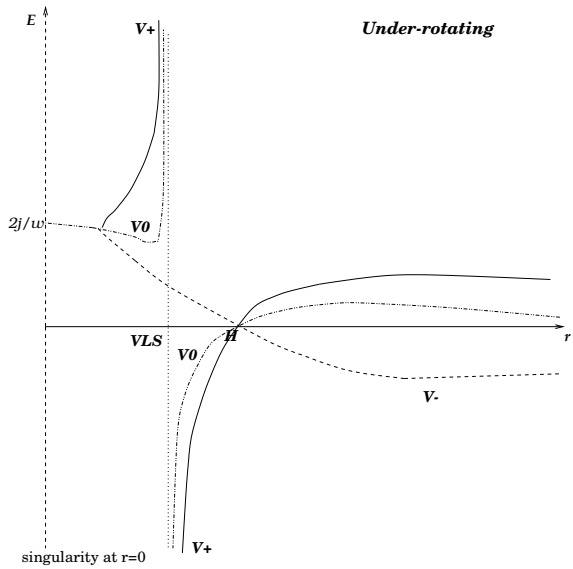,width=7.5cm}   
\caption{Effective potentials for a massive particle. Again the bifurcation point approaches the VLS as $m$ increases and the mass gap is created.}
\label{pot4}
\end{figure}

As we discussed in section 2, for a spacetime like the BMPV spacetime one can define generalized effective potentials due to the high degree of symmetry. This proves to be a powerful way of understanding motions in the spacetime. From (\ref{eqmr}) we write
\bequ
\left(\frac{dr}{d\lambda}\right)^2=\Delta_{L}(E-V^{-})(E-V^{+}),
\eequ
where
\bequ
V^{\pm}\equiv \frac{\Delta}{\Delta_{L}}\left[\frac{2\mu \omega j_{R}}{r^{4}}\pm \sqrt{\frac{4\mu^2\omega^2j_{R}^2}{r^8}+\left(m^{2}+\frac{4j^2}{r^2}\right)\Delta_{L}}\right].
\label{potentpm}
\eequ
Furthermore, we define another potential $V^0$ by
\bequ
\frac{dt}{d\lambda}=\frac{\Delta_{L}}{\Delta^2}\left[E-V^0\right]
\eequ
which will give us the relevant information on time travelling. We then read off from (\ref{dtdl}) 
\bequ
V^0\equiv \frac{2\mu \omega j_{R}}{r^4}\frac{\Delta}{\Delta_{L}}.
\eequ
Figures \ref{pot1} to \ref{pot4} elucidate the form of these potentials for equatorial motions. We show the $j_{R}$ positive case. The $j_{R}$ negative case is simply obtained by the inversion $E\rightarrow -E$, a reflection of the $PT$ invariance of axisymmetric spacetimes. For $r>r_L$ the allowed points in the graphs are either above or underneath both $V^{+}$ and $V^{-}$. For $r<r_L$ they are in between the two potentials.

\subsection{Analysis of geodesic motions}
Due to (\ref{jbound}) we can express $j_R=sj$, where $-1\le s\le 1$ is dimensionless and $j$ is always positive. Rewriting (\ref{eqmr}) as
\bequ
(\frac{dr}{d\lambda})^{2}=E^{2}\Delta_{L}-\Delta^{2}\left(m^{2}+\frac{4j^{2}}{r^{2}}\right)-\frac{4\mu \omega Esj\Delta}{r^{4}},
\label{radmot}
\eequ
the question of how ``far'' a freely falling particle can go is rephrased as ``when does the right hand side become negative ?". Clearly, a massless particle with no angular momentum following a geodesic will never penetrate the VLS. If we are in the over-rotating case it means that it will never enter the black hole region. A more interesting case occurs if the angular momentum of the particle is non-zero. We can have $j_R=0$ or $j_R\neq 0$. In the former case, geodesics will never enter the time machine, since the RHS of (\ref{radmot}) will become zero outside the VLS. But for the latter case, the last term (spin-orbit coupling) will contribute, and if the sign of $Es\omega\Delta$ is negative, there will be geodesic motions  going through the VLS. This contribution is maximal when $s=\pm 1$, which means that the motion is in the $\beta=\pi / 2$ plane, i.e. equatorial motions. Thus, particles ($E>0$) can only cross the VLS if $\Delta$ and $\omega s$ have different signs. This can be seen in figures \ref{pot1} and \ref{pot2}. Notice that the $PT$ counterpart of figure \ref{pot1} would show the over-rotating case where particles penetrate the VLS.

The above results can be readily understood if we recall the `Machian' dragging of inertial frames mentioned above. The condition we found, namely that $s\Delta$ needs to be negative (for particles and fixing $\omega$ to be positive), just means the geodesic motions need to have a spin \textit{opposite} to the dragging effects of the spacetime. Intuitively this is what we would expect. The particles have to be sent orbiting in the opposite direction to the rotation of the hole, so that the motion becomes \textit{sufficiently radial} as they approach the VLS, hence being able to enter it. Quantitatively this can be seen by writing (for equatorial motions):
\bequ
\barr{l}
\displaystyle{\left(\frac{dr}{d\lambda}\right)^{2}=E^{2}-m^{2}\Delta^{2}-\left(\frac{2j_R\Delta}{r}+\frac{\mu \omega E}{r^{3}}\right)^{2}  \ \ \ ,\ \ \ \frac{d\gamma}{d\lambda}=\frac{2}{\Delta r}\left[\frac{2j_R\Delta}{r}+\frac{\mu \omega E}{r^{3}}\right].}
\label{raddup}
\earr
\eequ 
Therefore, the radial velocity is highest, at any radial distance, when the (equatorial) motion is purely radial; in fact, if we want an orbit without rotation at some radial distance (since for geodesics it is not possible everywhere) we must compensate the dragging of frames by giving the particle some angular momentum. We may conclude that the necessary condition for penetrating the VLS ($j_R$ should be negative (positive) for the over-rotating (under-rotating) case), may be interpreted as a requirement on the motion to be sufficiently radial as it approaches $r_L$.

We now ask if freely falling particles can penetrate the black hole region.
For the over-rotating case it follows from (\ref{radmot}) that no geodesics will penetrate into the black hole region! This is quite striking, since according to our definition in section 2 it means that the manifold for $r>r_H$ is geodesically complete, and in fact the \textit{maximal analytical extension of the BMPV spacetime}. We will comment more on this below. For the under-rotating case, one has to deal with the usual phenomenon in black hole physics that the time coordinate $t$ (proper time of the observer at infinity) ``freezes" as one approaches the horizon. This follows from the divergence at $r_{H}$ in (\ref{dtdl}). But it can be seen that a freely falling test particle reaches the horizon for a finite value of a regular time parameter. Using the $v$ time coordinate as introduced in (\ref{vtime}), i.e. working in the patch $U_{in}$ (which removes the singularity at $r^{2}=\mu$ as long as $r_{L}\neq r_{H}$), the motion of an observer with no acceleration will obey:

\bequ
dv=\left[(1\pm 1)\frac{\Delta_{L}^{\frac{1}{2}}}{\Delta^{2}}\pm\left(\frac{m^{2}+\frac{4j^{2}}{r^{2}}}{2E^{2}\Delta_{L}^{\frac{1}{2}}}+2\left(\frac{\mu \omega j_R}{Er^{4}}\right)^{2}\Delta_{L}^{-\frac{3}{2}}\right)+\mathcal{O}(\Delta)\right]dr.
\eequ
The $-ve$ sign in the $\pm$ comes from orbits with $dr / d\lambda<0$, so that the divergent term drops out for ingoing trajectories, 
rendering a finite $v$ time for crossing the horizon. For rotating backgrounds, an additional divergence arises for the $\gamma$ coordinate. One can deal with it similarly, now using the $\gamma_{in}$ coordinate introduced in (\ref{gamin}). One then finds:

\bequ
d\gamma_{in}=\left[(1\pm 1)\frac{2\mu\omega}{\Delta r^4\Delta_L^{\frac{1}{2}}}\pm\left(\frac{4}{E\Delta_L^{\frac{1}{2}}r^2}\left(\frac{j_R-j_L \cos(\beta)}{\sin^{2}(\beta)}\right)-\frac{4\mu^2\omega^2 j_R}{r^4 E \Delta_L^{\frac{1}{2}}}\right)+\mathcal{O}(\Delta)\right]dr,
\eequ
with the negative sign coming from ingoing trajectories. 
Having clarified these usual artifact of the coordinates, we refer to figure \ref{pot2} which illustrates that particles with positive angular momentum will be able to enter the black hole region and also the time machine. Furthermore, except for the very special orbit with $E=2j / \omega$, all such orbits will bounce back at some radial distance $r_v$. Their motion will now be towards $r\rightarrow \infty$, and since they cannot go back to the same asymptotic region whence they originated, they must leave the patch $U_{in}$ and enter a new outgoing patch $U_{out}'$ containing a new asymptotically flat region of the complete BMPV spacetime. Therefore the global structure resembles the extreme Reissner-Nordstrom or the equatorial plane of the Kerr-Newman solution.

\subsection{Geodesic time travelling}

We now turn our attention to particles entering the time machine. The relevant figure is the $PT$ counterpart of figure \ref{pot1}. Clearly, all particles coming from infinity with enough energy to overcome the bump defined by the dashed line will penetrate the VLS. What happens to such particles? They cannot enter the black hole region and, although there are bound states with orbits in $r_{H}<r<r_{L}$, a particle coming from infinity will not fall into one of these states. Hence, all trajectories will come out again from the time machine region. So we are entitled to ask the question: is the time machine effective? Can a particle come out \textit{before} (in t time) it went in? From the figure we can see that the particle crosses the $V^{0}$ line, and hence will reach the region where $dt / d\lambda$ is negative. So, the overall $\Delta t$ for travelling from the VLS inwards and back to the VLS will have a negative contribution, and we now show that $\Delta t$ can indeed be negative.

Consider radial penetration in the VLS, i.e., the $\gamma$ component of the momentum 5-vector, $p^{\gamma}$, vanishes at $r_L$. Then, the ratio $j_R /E$ is fixed by the second equation in (\ref{raddup}). Furthermore, (\ref{radmot}) becomes
\bequ
\left(\frac{dr}{d\lambda}\right)^{2}=E^{2}\left[1-\left(\frac{\left(\frac{r_{L}}{r}\right)^{3}-\frac{r_{L}}{r}}{1-\left(\frac{r_{H}}{r_{L}}\right)^{2}}\right)^{2}\right]-m^{2}\Delta^{2}.
\label{drdlradpen}
\eequ
The radial velocity is therefore maximal at the VLS. For massless particles this maximum coincides with the value at infinity. The minimum will be at $r=\sqrt{3}r_{L}$ and, as long as $r_L$ is not too close to $r_H$, namely
\bequ
\frac{r_{L}}{r_{H}}\in \left]0,\left(1+\frac{8}{9\sqrt{3}}\right)^{-\frac{1}{2}}\right[ \bigcup \left] \left(1-\frac{8}{9\sqrt{3}}\right)^{-\frac{1}{2}},\infty\right[,
\label{interv}
\eequ
$\left(dr / d\lambda\right)^{2}$ will be a positive number everywhere outside the VLS, and therefore there will be geodesics coming from infinity and entering the time machine. In terms of the $PT$ counterpart of figure \ref{pot1}, 
(\ref{interv}) is the condition for the particle to overcome the bump defined by the dashed line, compatible with the $j_R/E$ ratio defined by radial penetration. Under such condition, (\ref{dtdl}) reads
\bequ
\frac{dt}{d\lambda}=\frac{Er^{2}}{\Delta^{2}(r_{L}^{2}-r_{H}^{2})}\left[\left(\frac{r_{L}}{r}\right)^{6}+ \left(\frac{r_{L}}{r}\right)^{2}-\left(\frac{r_{L}}{r}\right)^{8}-\left(\frac{r_{H}}{r}\right)^{2} \right],
\label{dtdlradpen}
\eequ
which is clearly positive on the VLS but is always negative (in fact diverges to $-\infty$) on the horizon. So, both $\left(dr / d\lambda\right)^{2}$ and $dt / d\lambda$ have a zero in $]r_{H},r_{L}[$, respectively at $r_{v},r_{t}$. To see that $r_{v}<r_{t}$ we drop the condition of radial penetration and instead parameterise $j_{R}$ as
\bequ
j_{R}=\frac{E}{2\mu \omega}r^{4}_{t}\frac{\Delta_{Lt}}{\Delta_t},
\eequ
where the subscript t corresponds to quantities computed at $r_t$. Then,
\bequ
\left(\frac{dr}{d\lambda}\right)^{2}_{t}=-m^{2}\Delta_{t}^{2}+E^{2}\left[1-\left(\frac{r_{t}}{r_{L}}\right)^{3}\right].
\eequ
So we conclude that the region $r<r_{t}$ is allowed if $r_{t}$ is inside the time machine, as shown in the figure. Moreover it is \textit{only} allowed if $r_{t}$ is inside the time machine. In other words, $dt/d\lambda$ can become negative outside the VLS if $j_R$ is positive. But that region is never visited by causal geodesics with $j_R>0$. This had to be the case, since we know that the light cones do not ``tip'' enough outside the VLS to allow time travel of timelike or null geodesics. 
More specifically, for a causal particle moving in the $t-\gamma$ plane, we can write the 5-momentum vector as $u=(d\gamma / d\lambda)\partial / \partial \gamma+(dt / d \lambda) \partial / \partial t$, and the causality requirement, $u^2\le 0$, reads:
\bequ
\left(\frac{2j_R}{r}-\frac{E}{\Delta}\left(1-\frac{\mu \omega}{r^3}\right)\right)\left(\frac{2j_R}{r}+\frac{E}{\Delta}\left(1+\frac{\mu \omega}{r^3}\right)\right)\le 0
\eequ
and it follows that for particles with positive $j_R$, outside the VLS, in the over-rotating case we must have $dt / d\lambda\ge 0$.

The time elapsed in a journey $r_{L}\rightarrow r_v\rightarrow r_{L}$ for a geodesic obeying (\ref{drdlradpen}) and (\ref{dtdlradpen}) is
\bequ
\Delta t=2r_L\int^{x_r}_1 \frac{1+x^4-x^6-a^{-2}}{\left(1-\left(\frac{x}{a}\right)^{2}\right)^2\left( x^2-\left(\frac{x}{a}\right)^2\right)\left(1-\left(\frac{x^3-x}{1-a^{-2}}\right)^2\right)^\frac{1}{2}}dx,
\eequ
where $a\equiv r_L / r_H$ and $x\equiv r_L / r$. Numerical integration shows that this is negative for any $a>1$, i.e. for the over-rotating case. As an example, for $a=2$, $\Delta t=-0.03r_L=-2*10^{-7}r_H (Km) seconds$, where we have used the usual four dimensional value for the velocity of light in the last equality, in order to provide some physical sensibility. In \cite{gibher} we will discuss more quantitatively geodesic time travel, in some simpler spacetimes.

\begin{figure}
\begin{picture}(0,0)(0,0)
\put(67,223){$\gamma$}
\end{picture}   
\centering\epsfig{file=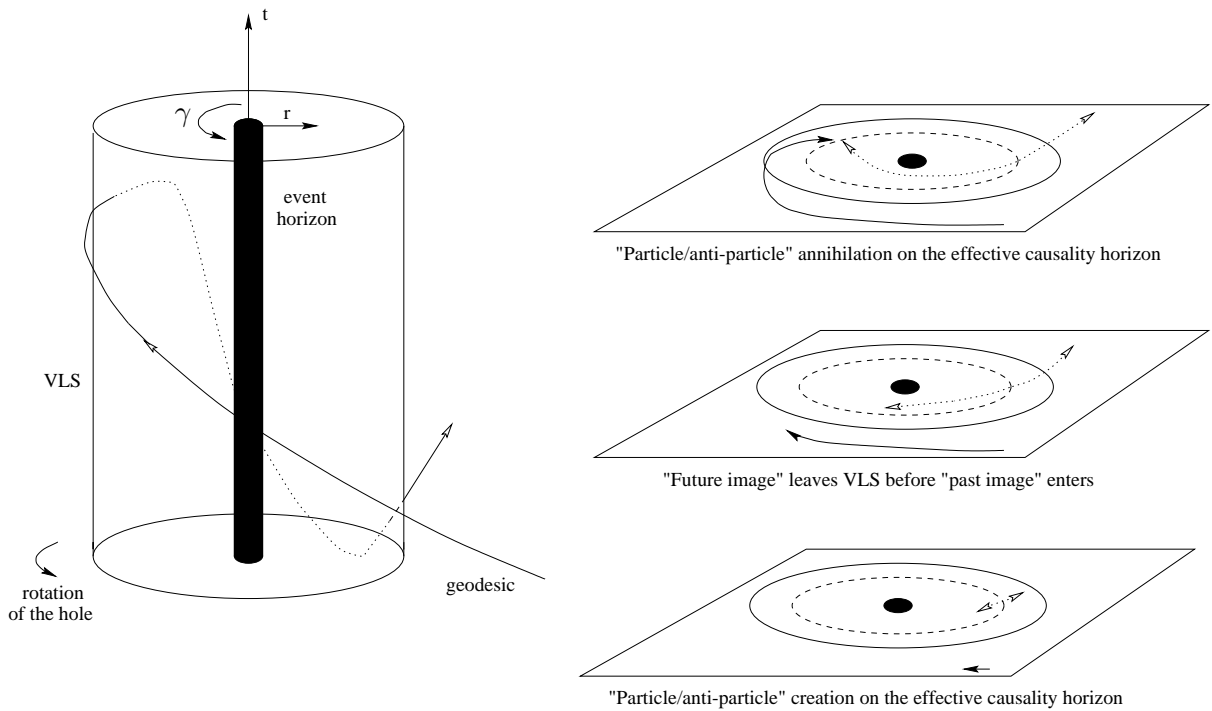,width=16cm}   
\caption{A picture of time travelling.}
\label{cyl}
\end{figure}

Figure \ref{cyl} gives an illustration of time travelling. The condition $j_R\omega<0$ for the particle to penetrate the VLS in the over-rotating case is illustrated. If the particle penetrates, it always reaches the ``effective causality horizon'', where $dt / d\lambda=0$. Then it moves downwards in our left diagram. Of course, the whole point is that the spacetime causal structure is described by light cones that tips substantially inside the VLS. As the rotation of the spacetime forces the particle to bounce back to increasing r, it leaves the causality horizon and also the VLS. But this happens before its past image actually entered the VLS. The right diagrams represent a constant t foliation of the left diagram. 

One might imagine trying to stop the particle from going in at the ``moment'' illustrated in the second figure on the RHS. The obvious paradox is then how could it emerge in the past? However, it should be borne in mind that the $t=constant$ hypersurfaces are not Cauchy surfaces and the function $t$ is not a global time function. Thus, one cannot specify arbitrary Cauchy data on a $t=constant$ hypersurface. In particular one is not free to choose Cauchy data corresponding to stopping the particle. An analogy would be to consider the problem in standard classical physics, electromagnetism say, where one would specify the initial data spread out over a non-Cauchy surface, like at different times. Inconsistencies can then easily be constructed. The difference is, obviously, that for the BMPV spacetimes, as in other manifolds with causal pathologies, there is no partial Cauchy surface. The Cauchy problem is always ill-defined.

\subsection {Carter-Penrose Diagrams}

A traditional way of exhibiting the causal structure
of static spherically symmetric black hole spacetimes
is by means of two-dimensional
Carter-Penrose (``CP'') diagrams. For such cases, the CP diagram encodes all the causal information about the spacetime. But our situation is more complicated and so we must look at extensions of the original examples to a more general construction. 

One direction in which one may generalise the concept is just
to regard a CP diagram
as a  two-dimensional totally geodesic submanifold of spacetime. This is what is done for the Kerr manifold, $M_{K}$. This solution possesses, in addition to the time translations, just one extra Killing field, $\bf m$, generating a circle action. Taking the quotient, $M_{K}/S^1$, would give a three-dimensional space, which is hard to visualize. As an alternative, one may find two-dimensional totally geodesic timelike submanifolds, corresponding to equatorial and polar motions. Then, two CP diagrams give some idea of the causal structure, but it is partial. For example, the equatorial CP diagram does not reveal the existence of CTC's inside the horizon, since these are to be found in the $S^1$ generated by $\bf m$. In the BMPV spacetime one can find totally geodesic submanifolds, in the same sense as for Kerr, but for our situation this does not work as well. The success of these diagrams for Kerr is to illustrate the difference with the charged spherically symmetric case, due to the ring singularity at $r=0$ and the new asymptotical region with negative $r$ coordinate. In the BMPV case, the singularity is point-like, so that diagram will not carry much new information. 

In the case of the Taub-NUT spacetime, the $U(2)$ isometry group does not act on two-dimensional orbits, but on three-dimensional orbits. These orbits are $S^1$ bundles over $S^2$, and together with the radial coordinate one may exhibit the spacetime as an $\Bbb R \times S^1$ bundle over $S^2$ \cite{hawell}. A CP diagram may be obtained from the restriction of the metric to the fibres. However, the orbits of the $S^1$ time translation action are \textit{not} orthogonal to the $S^2$ factor. In fact, the distribution $D \subset TM$ spanned by the radial and the time translation vectors is not integrable, and there is, unlike in the spherically symmetric case, no foliation by two-dimensional leaves, each of which may be identified with the CP diagram. The integrability property is  
sometimes referred to as {\it orthogonal transitivity}. The distribution is integrable if and only if the holonomy of the connection valued in the $G/H$ coset base space is trivial.

In our case, the orbits of the $SU(2)_L$ isometry subgroup correspond to $S^3$ and the spacetime may be though of as an $\Bbb R ^2$ bundle over $S^3$, the fibre directions being spanned by $\partial / \partial r$ and $\partial / \partial t$. If the rotation parameter $\omega$ vanishes, then we obtain  the usual Carter-Penrose
diagram for an extreme Reissner-Nordstrom spacetime, but if $\omega \ne 0$ things are more complicated.
The curvature of the bundle is non-vanishing and the orbits 
are not everywhere spacelike. Writing the BMPV metric as
\ben
ds^2 = -\frac{\Delta^2}{\Delta_L }dt^2 + 
{ dr^2 \over \Delta^2 } + 
{ r^2 \over 4} \left[
 (\sigma ^1)^2 
+ (\sigma ^2)^2  + 
\Delta  _L \left(\sigma ^3 -2 {\mu \omega \Delta \over r^ 4 \Delta_L}   dt\right)^2 \right],
\een
we see that the metric of the Carter-Penrose diagram is
\ben  
ds^2 = -\frac{\Delta^2}{\Delta_L }dt^2 + 
{ dr^2 \over \Delta^2 },
\label{metcp}
\een
and the $su(2)$-valued connection giving the holonomy of the 
distribution is
\ben
\left(0,0, -2 {\mu \omega \Delta \over r^4 \Delta_L} dt\right),
\een
giving for the $su(2)$-valued curvature
\ben
\left(0,0,4 {\mu  \omega(2r^6-3r^4r_H^2+r^6_L)\over r^{11}\Delta_L^2} dr \wedge dt\right) .
\een

As in the Taub-NUT case, the distribution spanned by $\frac{\partial}{\partial r}$ and $\frac{\partial}{\partial t}$ is not integrable, and the holonomy lies in a $U(1)$ subgroup. Thus, one does not have a foliation by CP diagrams. Nevertheless, the metric (\ref{metcp}) contains some information about the spacetime. However, \textit{it is not even everywhere Lorentzian}. It is positive definite inside the VLS. This is because, the $S^3$ orbits are spacelike outside the VLS but timelike inside it. This phenomenon does not occur in the Taub-NUT case.
 
\begin{figure}
\begin{picture}(0,0)(0,0)
\end{picture}   
\centering\epsfig{file=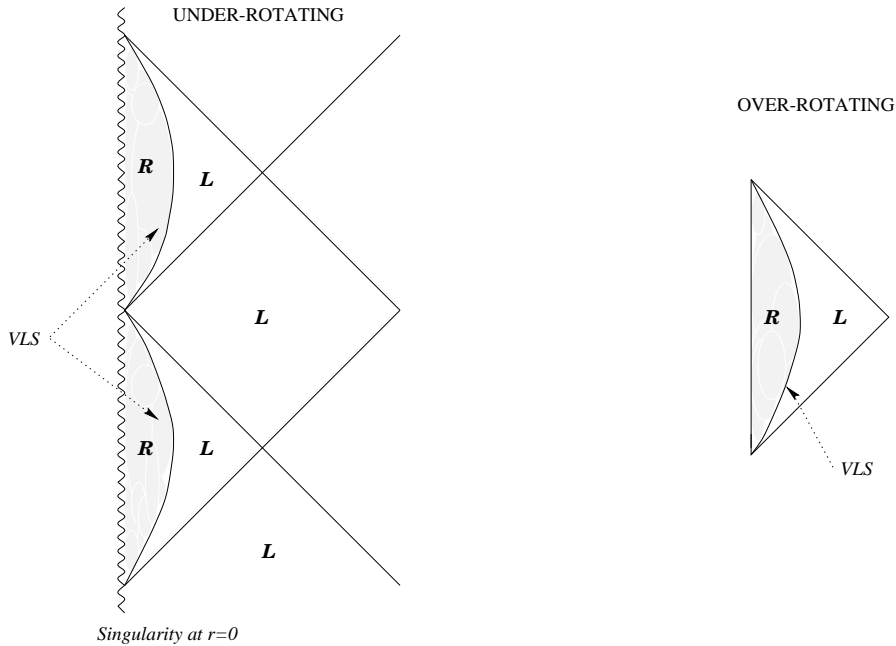,width=12cm}   
\caption{The Carter-Penrose diagrams for the BMPV spacetimes. The ``R'' and ``L'' denote the sections where the metric on the CP diagram is Riemannian and Lorentzian respectively.}
\label{CP}
\end{figure}

In figure \ref{CP} we show two Carter-Penrose diagrams for our spacetime. Although they provide some visualisation of the causal structure all of the above discussion should be borne in mind. One may think of them as corresponding to some totally geodesic submanifold, like the $\beta=\pi / 2$, $\alpha=constant$ submanifold.

\sect{Scalar field in BMPV background}
The Klein Gordon equation in this background is most easily computed using differential forms and the frames (\ref{frames}) and (\ref{framesinv}). Then, $\star d \star d\phi=m^{2}\phi$ reads:
\bequ
-\frac{\ddot{\phi}}{\Delta^{2}}+\frac{1}{r^{3}}\frac{\partial}{\partial r}\left(\Delta^{2} r^{3}\frac{\partial}{\partial r}\phi\right)+\frac{4}{r^{2}}\left({\bf L}_{1}^{2}+{\bf L}_{2}^{2}+\left({\bf L}_{3}-f\frac{\partial}{\partial t}\right)^{2}\right)\phi=m^{2}\phi.
\eequ
Clearly, eq. (\ref{hj}) is just the WKB approximation of the above equation.
Our ansatz to separate variables takes the form
\bequ
\phi=F(r)e^{-iE t}D^{j}_{j_{R},j_{L}},
\eequ
where the functions $D^{j}_{j_{R},j_{L}}$ are the Wigner D-functions (see for instance \cite{edmond}), which are simultaneous eigenfunctions of ${\bf L}_{3}$,${\bf R}_{3}$, ${\bf L}^{2}$ and ${\bf R}^{2}$. The eigenvalues are, respectively,
\bequ
\barr{c}
{\bf R}_{3}D^{j}_{j_R,j_L}=-ij_{L} D^{j}_{j_R,j_L}, \ \ \  \ \ \ {\bf L}_{3}D^{j}_{j_R,j_L}=ij_{R} D^{j}_{j_R,j_L},  
\\\\  {\bf L}^{2}D^{j}_{j_R,j_L}={\bf R}^{2}D^{j}_{j_R,j_L}=-j(j+1)D^{j}_{j_R,j_L}.
\earr
\eequ
With such an ansatz, the scalar field equation reduces to an ordinary differential equation (``ODE'') for the radial function $F(r)$:
\bequ
-\frac{\Delta^2}{r^3}\frac{d}{d r}\left(\Delta^{2}r^{3}\frac{d F}{d r}\right)=\left[E^2\Delta_{L}-\Delta^2\left(m^2+\frac{4j(j+1)}{r^2}\right)-\frac{4\mu \omega Ej_{R}\Delta}{r^4}\right]F.
\label{scar}
\eequ
For large $j$, we recover the RHS of (\ref{radmot}). Notice that for both cases the LHS is positive-definite. We can define a new variable, $x$, as to eliminate the first order term in this equation. This is achieved by the relation
\bequ
x=\frac{r_H^2}{r^2-r_H^2},
\eequ
the radial equation becoming
\bequ
\frac{d^2 F}{dx^2}=\left[A+\frac{B}{x}+\frac{C}{x^2}+\frac{D}{x^3}\right]F,
\label{scax}
\eequ
with coefficients
\bequ
\barr{c}
\displaystyle{A=\left(\frac{r_H E}{2}\right)^2\left[\left(\frac{r_L}{r_H}\right)^6 -1\right], \ \ \  \ \ \ B=r_H E\left[\left(\frac{r_L}{r_H}\right)^3 j_R -\frac{3}{4}E r_H\right],}
\\\\ \displaystyle{C=j(j+1)+\left(\frac{r_H}{2}\right)^2(m^2-3E^2), \ \ \  \ \ \ D=\left(\frac{r_H}{2}\right)^2 \left[m^2-E^2\right].}
\earr
\eequ
The general Klein-Gordon equation for the family of spacetimes found in \cite{cveyoum}, of which the BMPV black hole is a special case, was computed in \cite{cvelar}. However, the $x$ coordinate used therein fails for the BPS case, when the two horizons coalesce, and for such a special case some of the remarks made in \cite{cvelar} must be changed. So, whereas the equation therein possesses two regular singular points (horizons) and one irregular singular point (spatial infinity), (\ref{scax}) possesses two irregular singular points; at $x=0$, i.e. spatial infinity, and $x=\pm \infty$, corresponding to the degenerate horizon. Notice that $x\in ]-\infty,-1]$ parameterises the inside the black hole; $x\in [0,+\infty[$ parameterises the outside, and these are the only possible values for $x$. 

The singularity structure of (\ref{scax}) shows that the method for finding general solutions of second order linear ODE's fails. So we look for solutions of restricted validity.

\begin{description}

\item[Far horizon region ($r\gg r_H$):] It follows that (\ref{scar}) is approximated by
\bequ
\frac{d^2 F}{dr^2}+\frac{3}{r}\frac{dF}{dr}+\left[\epsilon^2-\frac{l(l+2)}{r^2}\right]F=0,
\label{farreg}
\eequ
where we have introduced $\epsilon^2=E^2-m^2$, $l=2j$, $l=0,1,2,...$. The ansatz $F=\chi / r$ and the rescaling $u=\epsilon r$ unveil the standard Bessel equation for $\chi$
\bequ
u^2\frac{d^2\chi}{du^2}+u\frac{d\chi}{du}+[u^2-(l+1)^2]\chi=0.
\eequ
Due to the integer order of this Bessel equation, the general solution 
is a linear combination of first and second kind (Neumann functions) Bessel functions,
\bequ
F=\frac{1}{u}\left[\alpha_+ J_{l+1}(u)+\alpha_{-}N_{l+1}(u)\right],
\label{bess}
\eequ
where $\alpha_{\pm}$ are constants. Since the asymptotic equation reveals little more than the dimensionality of spacetime, these results resemble other five dimensional computations \cite{malda,cvelpt}. 

Despite the fact that (\ref{bess}) is only valid for large $r$, it might correspond to large $u$ or small $u$, the latter case valid only for small $\epsilon$ ($\epsilon r_H\ll \epsilon r\ll 1$). Recall that for large and small values of the argument, the Bessel functions can be approximated by, respectively,
\bequ
\barr{l}
\displaystyle{J_n(x)\sim \sqrt{\frac{2}{\pi x}}\cos[x-(n+\frac{1}{2})\frac{\pi}{2}], \ \ \  \ \ \  J_n(x)\sim \frac{x^n}{2^n\Gamma (n+1)},}
\\\\ \displaystyle{N_n(x)\sim \sqrt{\frac{2}{\pi x}}\sin[x-(n+\frac{1}{2})\frac{\pi}{2}], \ \ \ \ \ \ N_n(x)\sim -\frac{\Gamma(n)}{\pi}\left(\frac{2}{x}\right)^{n}.}
\label{expansion}
\earr
\eequ
We remark that this expansion of $N_{n}(x)$ for small $x$ is only valid for $n>0$, but such will be our only case of interest. Using (\ref{expansion}), we get, for small $u$ (rewriting in terms of r),
\bequ
F\sim \alpha_{+} \frac{(r\epsilon)^l}{2^{l+1}\Gamma(l+2)}-\alpha_{-}\frac{2^{l+1}\Gamma(l+1)}{\pi(r\epsilon)^{l+2}},
\label{matchout}
\eequ
which we will use below to match with the near region solution. For large $u$,
\bequ
F\sim \sqrt{\frac{1}{2\pi u^3}}e^{i\frac{\pi}{2}(l+\frac{3}{2})}\left[e^{-iu}\left[\alpha_{+}+i\alpha_{-}\right]+e^{iu}e^{-i\pi(l+\frac{3}{2})}\left[\alpha_{+}-i\alpha_{-}\right]\right],
\eequ
the second term corresponding to an outgoing wave (scattered wave) and the first to an incoming wave (incident wave). We are interested in computing the absorption probability. If we knew the coefficients $\alpha_{\pm}$, we could immediately read off the scattering amplitude and from it deduce the absorption probability by unitarity:
\bequ
|\mathcal{A}|^2=1-\left|\frac{1-i\frac{\alpha_{-}}{\alpha_{+}}}{1+i\frac{\alpha_{-}}{\alpha_+}}\right|^{2}.
\label{absorption}
\eequ
These constants are constrained by regularity requirements. For flat space, for instance, (\ref{farreg}) would be valid all over spacetime; regularity at the origin would impose $\alpha_{-}$ to be zero, and we would get a scattering probability of one. In other words, ingoing and outgoing waves would differ only by a phase shift. There would be no absorption (since there is nothing to absorb) and no information loss. In our case we have to match the solution with a near region solution to learn more about those coefficients.

\item[Near horizon region ($r^2-r^2_H \ll \epsilon ^{-2}$):] 

In this region we can neglect both the $D / x^2$ term and the second term of the C coefficient in (\ref{scax}). It is clear from (\ref{scax}) that the behaviour of the solutions near the horizon ($x\rightarrow \infty$) will be $F\sim \exp{\sqrt{A}x}$, hence depending strongly on the sign of $A$, which distinguishes the under-rotating from the over-rotating case. The former possesses oscillating solutions, and hence there will be a flux near the horizon, and indeed some absorption by the black hole, but the latter possesses either a damping or a divergent solution on the horizon, which clearly indicates that there will be no flux, and hence no absorption by the black hole. This constitutes the \textit{quantum mechanical counterpart of the classical result of section 4}. Notice that is true for \textit{all} frequencies, including very high for which the WKB approximation is valid. Hence it contains as a particular case the result seen in the last section for the geodesics. For small frequencies we are also able to give quantitative details of the absorption process, which we now describe.  

By defining $2x\sqrt{A}=y$, the approximation of (\ref{scax}) in the near horizon region can then be put into the form 
\bequ
y^2\frac{d^2 F}{dy^2}=\left[\frac{y^2}{4}+\frac{B}{2\sqrt{A}}y+(j+\frac{1}{2})^2-\frac{1}{4}\right]F.
\label{whittaker}
\eequ
This is Whittaker's equation (see e.g. \cite{whitwat}), and the solutions are the Whittaker functions. These can be expressed in terms of other confluent hypergeometric functions, like the Kummer functions. Since $2j+1$ is an integer, the general solution can be written as a linear combination of first and second kind Kummer functions as follows:
\bequ
F=e^{-\frac{y}{2}}y^{j+1}\left[\beta_{+}M(j+1+\frac{B}{2\sqrt{A}},2j+2;y)+\beta_{-}U(j+1+\frac{B}{2\sqrt{A}},2j+2;y)\right],
\label{whitt}
\eequ
where $\beta_{\pm}$ are constants. For $A>0$ (over-rotating), $2\sqrt{A}x=Real(2\sqrt{A}x)>0$ as one approaches the horizon from the exterior. Hence, the asymptotic behaviour of $M$ and $U$ for sufficiently large $x$, i.e. sufficiently close to the horizon, gives rise to the following behaviour for F (in terms of the $x$ coordinate):
\bequ
F\sim \beta_+ \frac{\Gamma(2j+2)}{\Gamma(j+1+\frac{B}{2\sqrt{A}})} e^{x\sqrt{A}\left(1+\frac{B ln(x)}{2Ax}\right)}+\beta_- e^{-x\sqrt{A}\left(1+\frac{B ln(x)}{2Ax}\right)}.
\eequ

The leading behaviour is the one anticipated above. Requiring regularity of the wave function on the horizon implies $\beta_{+}=0$. 

For sufficiently small values of $\epsilon$, the near region solution will overlap the far region solution. In particular, for $\epsilon=0$, the near region solution actually covers the whole spacetime. If we choose $\epsilon$ sufficiently small, so that in the overlapping region 
\bequ
\epsilon^{-2}\gg r^2-r_H^2\gg2\sqrt{A}r_H^2,
\label{lowfreq}
\eequ
then $y$ will be small and we can approximate the solution (\ref{whitt}) by the first term. Rewriting in terms of r,
\bequ
F(r)\sim \beta_- \frac{(-1)^{l}r^{l}}{(l+1)\Gamma(\frac{B}{2\sqrt{A}}-j)\Gamma(j+1+\frac{B}{2\sqrt{A}})(2\sqrt{A}r^2_H)^{\frac{l}{2}}}.
\eequ

Matching with (\ref{matchout}) gives $\alpha_{-}=0$, and hence no absorption and a phase shift similar to the one obtained in flat space, in agreement with the flux method used above.

For the under-rotating case, the variable $y$ becomes purely imaginary, and care must be taken in computing the asymptotic expansion for the Whittaker functions. The most straightforward way is to replace the change to the variable $y$ by the change $t=\sqrt{|A|}x$. Then, we obtain Whittaker's equation for an imaginary argument, which is the Coulomb wave equation:
\bequ
t^2\frac{d^2 F}{dt^2}=\left[-t^2+\frac{B}{\sqrt{|A|}}t+j(j+1)\right]F.
\eequ
The solutions are the Coulomb wave functions \cite{handbook}:
\bequ
F=\beta_+F_j\left(\frac{B}{2\sqrt{|A|}},t\right)+\beta_-G_j\left(\frac{B}{2\sqrt{|A|}},t\right)
\label{coulomb}
\eequ
We know that near the horizon, the wave functions behave as $\exp{\pm it}$. As usual, we require the physical solution to be purely ingoing on the horizon, hence choose the $`+`$ sign in the exponential. This requirement singles out a specific linear combination of the two Coulomb wave functions. As $t\rightarrow \infty$, one finds the desired behaviour in the combination
\bequ
iF_j+G_j\sim e^{it(1-\frac{B}{2\sqrt{|A|}}\frac{ln(2t)}{t})}.
\eequ
Therefore we conclude that $\beta_{+}=i\beta_{-}$. Again, for small $\epsilon$ obeying (\ref{lowfreq}), we can approximate the solution (\ref{coulomb}) by the first term in the overlapping region:
\bequ
iF_j+G_j\sim iC_j \left(\frac{B}{2\sqrt{|A|}}\right)\frac{r_H^{l+2}\sqrt{|A|}^{l+1}}{r^{l+2}}+\frac{r^l}{(l+1)C_j \left(\frac{B}{2\sqrt{|A|}}\right)r_H^l \sqrt{|A|}^{j}},
\eequ
where 
\bequ
C_j(\eta)=\frac{2^j e^{-\frac{\pi \eta}{2}}|\Gamma(j+1+i\eta)|}{\Gamma(2j+2)}.
\eequ
Matching with (\ref{matchout}) can now be performed yielding
\bequ
\frac{\alpha_-}{\alpha_+}=-i\pi\frac{(\epsilon r_H)^{2l+2}\sqrt{|A|}^{l+1}e^{-\frac{\pi B}{2\sqrt{|A|}}}}{2^{l+2}\Gamma(l+1)^2\Gamma(l+2)^{2}} |\Gamma(1+\frac{l}{2}+i\frac{B}{\sqrt{|A|}})|^2.
\eequ
Since the coefficient of $-i$ is real and positive, it can be seen from (\ref{absorption}) that the scattering probability is never greater than one; in other words, super-radiant scattering does not occur for the BMPV black hole, in contrast with four dimensional Kerr-Newman. This was to be expected as a consequence of the preserved supersymmetry of this background. Notice also the $l$ factor in the exponent of $\epsilon$ which means that the leading order term for the absorption probability comes from the s-wave. 
\end{description}

Using (\ref{absorption}) we get in the limit of small $\epsilon$:
\bequ
|\mathcal{A}|^{2}=\pi\frac{(\epsilon r_H)^{2l+2}\sqrt{|A|}^{l+1}e^{-\frac{\pi B}{2\sqrt{|A|}}}}{2^{l}\Gamma(l+1)^2\Gamma(l+2)^{2}} |\Gamma(1+\frac{l}{2}+i\frac{B}{\sqrt{|A|}})|^2
\eequ
The absorption cross section is obtained by multiplying the absorption probability by the appropriate phase space factor \cite{gubser}. Of particular interest is the partial cross section for the s-wave, which approximates the total cross-section and gives
\bequ
\sigma_{abs}^{l=0}=|\mathcal{A}|^{2}\frac{4\pi}{\epsilon^3}=\frac{4\pi^3}{\epsilon}\frac{r_H^{2}B}{e^{\frac{\pi B}{\sqrt{|A|}}}-1}.
\eequ
Since $\pi B / \sqrt{|A|}$ is small, the cross-section for small frequencies reduces to
\bequ
\sigma_{abs} \simeq \sigma_{abs}^{l=0}=\frac{E}{\epsilon}Area_{BMPV}.
\label{cross}
\eequ
For the massless wave case this generalises the result of \cite{das} for a rotating black hole: \textit{the absorption cross section for the s-wave is still the area of the black hole}, despite the loss of spherical symmetry of the horizon. This is in agreement with the behaviour found a long time ago for the four dimensional Kerr black hole \cite{page} and with the result for non-extreme black holes of the five dimensional family found in \cite{cvelar}. 

As was noted in the original paper \cite{BMPV}, and on which we will comment in the next section, the classical entropy formula for the over-rotating case gives an imaginary quantity and hence becomes senseless. Now, we have seen that in such a case the absorption probabiliy and cross section are zero. Since (\ref{cross}) is telling us that $\sigma_{abs}$ is a measure of the area in the under-rotating case, these two results seem to indicate that the effective area and entropy of the BMPV over-rotating black hole should be considered to be zero. Notice, however, that the derivation of (\ref{cross}) breaks down at the critical point.

The above calculation was only valid for small $\epsilon$, which for the massless case means small frequency of the wave. For very high frequencies, one approaches the particle limit, so that the WKB approximation is justified; in other words, the scattering problem is reduced to the computation of the ``radius of capture", $R_c$, for the classical geodesic motions. In general, the results then obtained differ from the ones at small frequency, that is, the absorption cross section has a frequency dependence. For the Schwarzchild D-dimensional black hole, the absorption cross section in the small and high frequency limits are, respectively (for massless spinless particles),
\bequ
\sigma^{low}=\frac{2\pi^{\frac{D-1}{2}}}{\Gamma(\frac{D-1}{2})}(r_H)^{D-2}, \ \ \ \ \ \ \sigma^{high}=\frac{2\pi^{\frac{D-1}{2}}}{\Gamma(\frac{D-1}{2})}\left(\frac{D-1}{2}\right)^{\frac{D-2}{D-3}}(r_H)^{D-2}.
\eequ 
These are areas of spheres; in four dimensions they have radii $r_H=2M$ and $3M$ respectively. The latter is the well known radius for the unstable circular orbit ``sitting on top" of the potential for null orbits in the Schwarzchild spacetime. For the extreme Reissner-Nordstrom solution, the formula for low frequencies is the same, but for high frequencies becomes
\bequ
\sigma^{high}=\frac{2\pi^{\frac{D-1}{2}}}{\Gamma(\frac{D-1}{2})}\left(D-2 \right)^{\frac{D-2}{D-3}}(r_H)^{D-2}.
\eequ
For the four dimensional case this corresponds to the area of a sphere of radius $2M=2r_H$ and for the five dimensional case a sphere of radius $\sqrt{3}r_H$. In the BMPV case, one can compute $R_c$ by using the potentials (\ref{potentpm}). Specializing to equatorial motions, and dealing with the particle case (the antiparticle is analogous), it then follows that one has to consider two cases: particles with an angular momentum component $j_R$ with the same or opposite sign to $\omega$, the black hole rotation parameter. It is straightforward to verify that $R_c$ will be given by the largest solution of the cubic polynomial equation
\bequ
r^3-3rr_H^2\pm 2r^3_L=0,
\label{poly}
\eequ
with `$+$' and `$-$' corresponding respectively to $j_R \omega$ positive or negative. The trigonometric solution is more useful than the usual Cardano's solution and yields
\bequ
R_c=2r_H\cos\left[\frac{1}{3}\arccos\left(\mp\left(\frac{r_L}{r_H}\right)^3\right)\right],
\eequ
respectively. Thus, for $r_L=0$, i.e., no rotation, we get $R_c=\sqrt{3}r_H$, in agreement with the above remarks for an extreme Reissner Nordstrom geometry. Naturally, this case is not sensitive to the angular momentum of the particle. Turning on the rotation, however, the behaviour depends on the sign of $j_R \omega$. As $r_L$ increases from 0 to $r_H$,
\bequ
\barr{l}
R_c \   decreases \ from \ r=\sqrt{3}r_H \ to \ r_{H} \ for \ j_R\omega>0,
\\\\
R_c \   increases \ from \ r=\sqrt{3}r_H \ to \ 2r_{H} \ for \ j_R\omega<0.
\earr
\eequ
The result is readily interpreted from the analysis of section 4. The dragging effect facilitates the fall into the black hole of particles with the opposite angular momentum from that of the hole ($j_R\omega<0$). Hence the radius of capture becomes larger. The opposite reasoning holds when particle and hole have the same sign of angular momentum. It then follows that the absorption cross section for high frequency scalar waves obeying the relation $j=\pm j_R$, in the under-rotating BMPV spacetime is
\bequ
\sigma^{high}=2\pi^2\sqrt{\left(2r_H\cos\left[\frac{1}{3}\arccos\left(\mp\left(\frac{r_L}{r_H}\right)^3\right)\right]\right)^6-r_L^6}.
\eequ
In terms of Figure \ref{pot2}, the radius of capture corresponds to the extreme of the potentials outside the horizon: for $j_R \omega>0$ the maximum of $V^+$; for $j_R \omega<0$, the minimum of $V^-$ (which is the maximum of $V^+$ in the $PT$ counterpart of figure \ref{pot2}). Clearly, this analysis is only valid for the under-rotating case. For the over-rotating we know that the absorption cross section is zero. In fact, for $j_R\omega>0$, we see from figure \ref{pot1} that $V^{+}$ has no ``bump'' outside the horizon, which corresponds to the non existence of a positive real zero for equation (\ref{poly}). For $j_R\omega<0$, we would be looking at the $PT$ counterpart of figure \ref{pot1}. Then, there is a minimum for $V^-$ outside the horizon, but this cannot be seen as measuring any effective region for absorption since no particles will be captured by the black hole.

\sect{Thermodynamics - The particle/tachyon analogy}
The usefulness of the BMPV spacetime for ``practical" time travel seems to be extremely limited. Firstly because it is a five dimensional spacetime. Secondly, because as for an ordinary extreme Reissner-Nordstrom black hole, the third law of thermodynamics seems to prevent the formation of such a BPS spacetime from a non-extreme one, by means of a finite process. The inability to form the infinite throat (which is also to be found in the BMPV spacetime) under any finite process is one way to understand it. This would represent the classical argument to explain why we could not construct the stable naked time machine we have described above. One could also try to explore quantum inconsistencies in forming such manifold from a causally well behaved initial configuration. However, arguments along the lines of \cite{cpc} do not seem to apply, due to the singular nature of the under-rotating spacetime, which seems to be an inevitable passage point in order to achieve the over-rotating one. Other lines of reasoning, as in \cite{cass}, find no application in our case either, since our scalar field computation seems to show that scalar eigenfunctions would exist even for the naked time machine spacetime. We will not persue this investigation in this paper, but, nevertheless, we would like to give a simple thermodynamical argument presenting some evidence that the passage from the under to over-rotating spacetime is forbidden. It relies on a simple analogy that can be drawn between the behaviour of the under-rotating or over-rotating BMPV black holes and, respectively, an ordinary particle and a tachyon.

For an ordinary relativistic particle, the energy relation $E=\sqrt{m^2+{\bf p}^2}$ yields a ``first law of thermodynamics" of the type:
\bequ
dE={\bf v}.d{\bf p}+\left(\frac{1}{4m_i\gamma}\right)\frac{dA}{8\pi}.
\eequ
We have defined the ``area'' as $A=16\pi m_i^2$, where $m_i$ is the \textit{irreducible} mass, which coincides with the rest mass for the particle, and $\gamma$ is the usual Lorentz contraction factor. The 3-velocity and the 3-momentum appear as conjugate variables and we can define the ``susceptibility'' $\chi$ by 
\bequ
\chi=\left(\frac{\partial^2 E}{\partial p^2}\right)_{m_i}=\left(\frac{\partial v}{\partial p}\right)_{m_i}=\frac{m^2}{(m^2+{\bf p}^2)^{\frac{3}{2}}},
\eequ
which vanishes as $p\rightarrow \infty$. The susceptibility measures how receptive the system is to change the velocity due to a change in momentum, under a reversible process. The fact that it vanishes just means that the velocity approaches asymptotically the speed of light, as can be seen by the standard relation:
\bequ
{\bf v}=\frac{\bf p}{\sqrt{m^2+p^2}}.
\eequ
Notice that unlike the velocity, the energy is not bounded and diverges as we take $p\rightarrow \infty$. But both energy and velocity increase as we increase $p$. For the latter this is encoded in the positivity of $\chi$.

The mechanical relations for a tachyon can be obtained by replacing $m\rightarrow im$. The modulus of the 3-momentum becomes constrained to take values in $]m,+\infty[$, and the susceptibility becomes negative. Hence, as we give more momentum to the tachyon, it \textit{slows down}, approaching asymptotically the velocity of light (from above) as $p\rightarrow \infty$. On the other hand, as it approaches the minimum value for momentum its velocity diverges. Notice, however, that just as for an ordinary particle, the energy always increases with $p$.

Now consider the Kerr black hole. The area of the intersection of the future event horizon with some partial Cauchy surface is
\bequ
A=8\pi(M^2+\sqrt{M^4-J^2}).
\label{areakerr}
\eequ
We define the irreducible mass as before, $A=16\pi m_i^2$, which now coincides with the Schwarzchild mass. For a reversible process we then find
\bequ
dM=\frac{JdJ}{4m_i^2\sqrt{m^2_i+\frac{J^2}{4m^2}_i}},
\eequ
where the coefficient of $dJ$ is, of course, the angular velocity of the horizon, $\Omega_H$. As for the particle case, the susceptibility, which is now $\left(\partial \Omega_H / \partial J\right)_{m_i}$, goes to zero as $J\rightarrow \infty$, and is always positive. Thus $\Omega_H$ tends asymptotically to a finite value, namely $(2m_i)^{-1}$. But differently from the particle case, there is now a bound on the possible values of $J$; it must be in $]0,2m^2_i]$. For higher $J$, (\ref{areakerr}) with a fixed $m_i$ has no real solution for $M$. Therefore, the maximum allowed value for $\Omega_H$ is slightly smaller than the asymptotic value: $(2\sqrt{2}m_i)^{-1}$. This corresponds to the angular velocity of an extremal Kerr black hole.

For the BMPV black hole, the area of the future event horizon is
\bequ
A=4\pi\sqrt{\frac{16M^3}{27\pi}-J^2_R},
\label{areabmpv}
\eequ
where $M,J_R$ are the spacetime mass and angular momentum, related with the metric parameters by (\ref{ADM}) and (\ref{ADMJ}). The irreducible mass for a given area is defined by (\ref{areabmpv}) with $J_R=0$, as to coincide with the mass for a five dimensional Schwarzchild black hole. The first law for a reversible process takes the form
\bequ
dM=\frac{9\pi}{8}\frac{J_R}{\left(m_{irr}^3 + \frac{27\pi}{16}J_R^2\right)^{\frac{2}{3}}}dJ_R,
\label{omegabmpv}
\eequ
where the coefficient of $dJ_R$ should be interpreted as $\Omega$, an averaged angular velocity of the spacetime and not the angular velocity of the horizon, which in fact is zero. (We remark that unlike the Kerr metric, the BMPV spacetime is \textit{not} rigidly rotating.) Defining $\Omega_c$ as the angular velocity for a critically rotating ($16M^3=27\pi J_R^2$) spacetime, we see that 
\bequ
\frac{\Omega}{\Omega_c}=\left(1+\frac{16m_{irr}^3}{27\pi J^2_R}\right)^{-\frac{2}{3}}
\label{omegarel}
\eequ
The relative susceptibility is
\bequ
\chi_{rel}=\left(\frac{\partial \frac{\Omega}{\Omega_C}}{\partial |J_R|}\right)_{m_i}=\frac{4^3 m_{irr}^3}{3^4\pi |J_R|}\left[1+ \frac{m_{irr}^3}{\frac{27\pi J_r^2}{16}}\right].
\label{suscep}
\eequ

Let us analyse the two cases:
\begin{description}
\item[Under-rotating case ($|J_R|< \sqrt{16/(27\pi)}M^{\frac{3}{2}}$) -]
In contrast with the Kerr case we now can give as much angular momentum to the spacetime as we wish, keeping the area fixed. By doing so we approach asymptotically the critical angular velocity from below, $\Omega/\Omega_c { \stackrel {J_R\rightarrow \infty} {\longrightarrow} 1} $. Therefore, an under-rotating black hole cannot become an over-rotating black hole by means of a reversible process, in analogy to the impossibility of a particle to overtake the velocity of light. Moreover, the susceptibility is indeed always positive.

\item[Over-rotating case ($|J_R|> \sqrt{16/(27\pi)}M^{\frac{3}{2}}$) -]
The immediate problem seems to be that the area becomes imaginary, and so does the entropy. But this is just a manifestation of the fact that a t=constant section of the ``horizon'' is no longer a spacelike hypersurface. In fact, the usual derivation of thermodynamical results breaks down, since the spacetime has a VLS outside the horizon. But let us still analyse a process of constant area (which is now an imaginary number) as being a reversible process. The values of $\Omega/\Omega_c$ and $\chi_{rel}$ are given by (\ref{omegarel}) and (\ref{suscep}) by taking $m_{irr}^3$ to be negative. As for a tachyon, the allowed values for $J_R$ are bounded below, $16/(27\pi)m_{irr}^3<J_R^2<\infty$. Hence, in complete analogy with the tachyon case, $\Omega/\Omega_c$ is always bigger than one, but it decreases, approaching one asymptotically, as we increase the angular momentum of the spacetime. 
\end{description}

\section{Conclusions}
In this paper we have discussed several issues concerned with the five dimensional solution found by Breckenridge, Myers, Peet and Vafa. The new features of this solution are the co-existence of supersymmetry and a non-trivial angular momentum for the spacetime. Furthermore, to the best of our knowledge, it provides the first example of a supersymmetric solution with trivial topology describing, for some values of the parameters, a naked, totally non-singular time machine. This led us to review the notions of time functions, time coordinates and develop a general framework for stationary spacetimes admitting a $S^1$ action, which we did in the first section. We furthermore reviewed the notion of gravito-magnetism and introduced the notion of the Sagnac connection. Such ideas will been more deeply scrutinised in \cite{gibher}. There we show, in some well-known examples of rotating spacetimes, how the computation of the Landau levels on ${\Bbb R}^2$, ${\Bbb H}^2$ or $S^2$ is equivalent to the the study of the Klein-Gordon field in particular spacetimes. These spacetimes are such that ${\Bbb R}^2$, ${\Bbb H}^2$ or $S^2$ arise as spacelike hypersurfaces orthogonal to the timelike killing vector field, and the Sagnac connection provides the magnetic field on the Riemann spaces. We also note that for the cases where spacetime can be seen as a Lie group, a simple group theoretical argument exists reproducing the same computation.

In order to understand thoroughly the structure of the BMPV spacetime we studied the geodesic equations, which can be completely separated due to the large isometry group. For this purpose, the effective potentials were exhibited. The existence of a reducible St\"ackel Killing tensor was discussed in section 3, and the analysis of geodesic motions was done in section 4. Two interesting features arise at this classical level: the dragging effects associated with the rotation and the geodesic time travel effects. The latter are allowed by the fact that the orbits of $U(1)$ subgroups of the isometry group become timelike. It is non-trivial, however, that even inertial observers can experience time travel. In fact, the condition that freely falling motions can penetrate the VLS induces an effect against geodesic time travel. Such condition is interpreted in terms of the dragging effects. We use this example to state clearly the problem associated with most spacetimes with causality violation: The inexistence of a Cauchy surface does not allow a well defined Cauchy problem. Interpreting time travel in terms of particle/antiparticle creation or annihilation events, as in figure \ref{cyl}, seems quite appealing. But it is not quite clear as yet how far one can push such interpretation.

The global structure can be read from the study of the geodesics. We have drawn the Carter-Penrose diagrams, but emphasised the subtleties and the limitations of such devices for non spherically symmetric manifolds. Particularly interesting is the behaviour that an over-rotating spacetime is found to possess. From the viewpoint of causal geodesic motions, a timelike boundary for the spacetime is created at $r=r_H$. One is then tempted to think along classical lines: ``the strong rotation creates enormous centrifugal forces which prevent fall into the black hole''. Such an argument misses, however, two points. Firstly, what one classically thinks of as centrifugal forces is encoded in the $j^2/r^2$ term present in the potentials, whereas the relativistic corrections, namely the term coming from the Sagnac connection, are what actually supports the `repulson' effect. Secondly, it is non-trivial (and related to the previous point) that BMPV spacetimes with this repulson effect coincide with the ones possessing CTC's outside $r=r_H$. The issue of understanding the behaviour of charged causal particles (and waves) in the BMPV spacetime will be addressed in \cite{gibher}.

The `repulson-like' behaviour for the over-rotating manifold is confirmed in section 5 by studying the Klein-Gordon field on the BMPV background. Furthermore, it is shown not to be a peculiar behaviour of high frequency waves, but in fact present for all frequencies. The study of an uncharged scalar field on the under-rotating background follows a standard prescription. The outcome generalises for BMPV spacetimes the well known result that the absorption cross section for low frequencies coincides with the area of the hole. For high frequencies we use the WKB approximation to derive the cross section. As one should expect, $\sigma_{abs}$ depends on the angular momentum of the wave. This is due to the spacetime's dragging effects.

The well known fact that one cannot spin up a non-extreme Kerr-Newman black hole to extremality by a finite reversible process finds a similar counterpart in the BMPV spacetime, as shown in section 6. Of course one cannot regard this as conclusive evidence towards the impossibility of creating the naked time machine from the original black hole causal spacetime. The analysis is quite simple minded, since even to maintain the black hole in the BMPV family, the changes in mass and angular momentum should be accompanied by changes in the charges. A more thorough study must therefore be performed. But it is tempting to think that such should be the case. A strong motivation arises from the fact that the CFT description of the D-brane system possesses non-unitary states for such naked time machine \cite{BMPV}. Hence such transition would involve loss of unitarity.

The theories of gravity available at the moment possess solutions which seem to allow displacement both forwards and backwards in a timelike direction. This is true not only for general relativity or supergravity but also for non-standard gravity theories (for a large list of references see \cite{post}). Classically, there seems to be no general argument why one should discard such solutions. Quantum mechanics seems to be more promissing as a `chronology protector'. The example discussed in this paper shows that string theory, the most successful quantum theory of gravity so far, seems to have the potential to shed some light on causality violations issues, namely by exploring the relation between macroscopic causality and microscopic unitarity. But it is not clear as yet, what the general statement, if any, will be.

\section*{Acknowledgments}

C.H. is supported by FCT (Portugal) through grant no. PRAXIS XXI/BD/13384/97.

\end{document}